\documentclass[amsmath,amssymb,twocolumn]{aastex63}
\usepackage{color}
\usepackage{amsmath}
 \usepackage{ulem}

\begin{document}


\title{Electromagnetic counterparts of binary neutron star mergers leading to a strongly magnetized long-lived remnant neutron star} 


\author{Kyohei Kawaguchi}
\affil{Institute for Cosmic Ray Research, The University of Tokyo, 5-1-5 Kashiwanoha, Kashiwa, Chiba 277-8582, Japan}
\affiliation{Center for Gravitational Physics,
 Yukawa Institute for Theoretical Physics, 
Kyoto University, Kyoto, 606-8502, Japan} 
\author{Sho Fujibayashi}
\affil{Max Planck Institute for Gravitational Physics (Albert Einstein Institute), Am M\"{u}hlenberg 1, Potsdam-Golm, 14476, Germany}
\author{Kenta Hotokezaka}
\affil{Research Center for the Early Universe, Graduate School of Science, University of Tokyo, Bunkyo-ku, Tokyo 113-0033, Japan}
\author{Masaru Shibata}
\affil{Max Planck Institute for Gravitational Physics (Albert Einstein Institute), Am M\"{u}hlenberg 1, Potsdam-Golm, 14476, Germany}
\affiliation{Center for Gravitational Physics,
 Yukawa Institute for Theoretical Physics, 
Kyoto University, Kyoto, 606-8502, Japan} 
\and
\author{Shinya Wanajo}
\affil{Max Planck Institute for Gravitational Physics (Albert Einstein Institute), Am M\"{u}hlenberg 1, Potsdam-Golm, 14476, Germany}
\affiliation{Interdisciplinary Theoretical and Mathematical Science (iTHEMS) Research Group, RIKEN, Wako, Saitama, 351-0198, Japan}

\newcommand{\rednote}[1]{{\color{red} (#1)}}
\newcommand{\KK}[1]{{\color{red} #1}}
\newcommand{\addms}[1]{{\color{blue} #1}}
\newcommand{\SF}[1]{{\color{magenta} #1}}
\newcommand{\SW}[1]{{\color{cyan} #1}}
\newcommand{\KH}[1]{{\color{orange} #1}}

\begin{abstract}
We explore the electromagnetic counterparts that will associate with binary neutron star mergers for the case that remnant massive neutron stars survive for $\agt 0.5$\,s after the merger. For this study, we employ the outflow profiles obtained by long-term general-relativistic neutrino-radiation magneto-hydrodynamics simulations with a mean field dynamo effect. We show that a synchrotron afterglow with high luminosity can be associated with the merger event if the magnetic fields of the remnant neutron stars are significantly amplified by the dynamo effect. We also perform a radiative transfer calculation for kilonovae and find that for the highly amplified magnetic field cases, the kilonovae can be bright in the early epoch, while it shows the optical emission rapid declining in a few days and the long-lasting ($\sim 10\,{\rm d}$) emission very bright in the near-infrared wavelength. All these features have not been found in GW170817, indicating that the merger remnant neutron star formed in GW170817 might have collapsed to a black hole within several hundreds ms or magnetic-field amplification might be a minor effect. 
\end{abstract}

\section{Introduction}\label{sec:intro}
The merger of neutron stars (NSs) is currently one of the most interesting multi-messenger phenomena in high-energy astrophysics, in which physical processes in extreme (strongly self-gravitating, high-density, and high-temperature) environments are realized. During the inspiral motion just before the merger, gravitational waves reflecting the mass, spin, and internal structure of the NSs are emitted. At the onset of the merger, the NS matter is ejected by tidal disruption and collisional shock heating~\cite[e.g.,][]{Rosswog:1998hy,Ruffert:2001gf,Hotokezaka:2012ze}. After the binary merger, a massive neutron star (MNS) or a black hole (BH) surrounded by a strongly magnetized hot and dense accretion torus is formed~\citep{Price:2006fi,Kiuchi:2017zzg}. The accretion torus is considered to launch a relativistic jet and outflows by magnetic pressure and tension, viscous heating due to magneto-hydrodynamical turbulence, and neutrino irradiation. In addition, the $r$-process nucleosynthesis of heavy elements is expected to proceed in the neutron-rich ejecta~\citep{Lattimer:1974slx,Eichler:1989ve,Freiburghaus1999a,Cowan:2019pkx}. In such a situation, weak interaction processes play an important role in determining the thermodynamic properties of the merger remnants, the post-merger environment, and the abundance of the elements synthesized in the ejecta~\cite[e.g.,][]{Metzger:2010sy,Goriely:2010bm,Wanajo:2014wha,Just:2014fka,Sekiguchi:2015dma,Sekiguchi:2016bjd,Radice:2016dwd,Miller:2019dpt,Fujibayashi:2017puw,Fujibayashi:2020qda,Fujibayashi:2020jfr,Fujibayashi:2020dvr,Foucart:2020qjb,Just:2021cls}.

The observations of gravitational waves from an NS-NS merger (GW170817;~\citealt{TheLIGOScientific:2017qsa}) and its multi-wavelength electromagnetic (EM) counterparts ~\citep{GBM:2017lvd} have demonstrated that such observations provide an invaluable opportunity to study fundamental physics under extreme conditions. For example, gravitational waves detected enable us to infer the degree of the NS tidal deformation in close orbits, leading to constraining the NS equation of state (EOS)~\citep{TheLIGOScientific:2017qsa,De:2018uhw}. Observations in the optical and near-infrared (NIR) wavelengths ~\citep{Andreoni:2017ppd,Arcavi:2017xiz,Coulter:2017wya,Cowperthwaite:2017dyu,Diaz:2017uch,Drout:2017ijr,
Evans:2017mmy,Hu:2017tlb,Valenti:2017ngx,Kasliwal:2017ngb,Lipunov:2017dwd,Pian:2017gtc,Pozanenko:2017jrn,Smartt:2017fuw,Tanvir:2017pws,Troja:2017nqp,Utsumi:2017cti} are consistent with the theoretical predictions for a kilonova; the emission powered by the radioactive decay heating of $r$-process elements~\citep{Li:1998bw,Kulkarni:2005jw,Metzger:2010sy,Kasen:2013xka,Tanaka:2013ana}. The rapid color evolution and spectra strongly suggest that indeed $r$-process elements have been synthesized in the ejecta~\citep[e.g.,][]{Cowperthwaite:2017dyu,Kasen:2017sxr,Kasliwal:2017ngb,Perego:2017wtu,Tanaka:2017qxj,Villar:2017wcc,Rosswog:2017sdn,Kawaguchi:2018ptg}. 

A synchrotron afterglow was discovered in the radio, optical, and X-ray bands~\citep{Hallinan2017Sci,Haggard2017ApJ,Margutti2017ApJ,Troja2017Natur,Lyman2018NatAs}. A superluminal motion of the radio source reveals that a narrowly collimated relativistic jet is produced in GW170817 \citep{Mooley2018Natur}.
In addition to the jet afterglow, a long-term synchrotron flare is expected to arise from the merger ejecta on time scales of $\sim 1$--$10^4$ years~\citep{Nakar2011Natur,Hotokezaka:2015eja,Hotokezaka:2018gmo,Margalit2020MNRAS}. Interestingly, the  X-ray flux observed around 3.5 years after GW170817   exceeds the expected flux of the jet afterglow, suggesting that the ejecta afterglow starts to dominate over the jet component \citep{Hajela:2021faz,Troja2022MNRAS}. 
However, the origin of the X-ray excess is sill under debate because the radio flux is still consistent with the prediction of the jet afterglow \citep{Balasubramanian2021ApJ}.

There are  a variety of open questions associated with GW170817. One of the most important questions is whether, and if so, when the remnant NS has gravitationally collapsed into a BH. This question is connected to the determination of the maximum mass of the NS, which places a  constraint on the NS EOS~\citep[e.g.,][]{Margalit:2017dij,Rezzolla:2017aly,Shibata:2019ctb}. The amount of the ejecta mass and its ejection mechanism are also suggestive for understanding the underlying physics~\citep[e.g.,][]{Bauswein:2017vtn,Coughlin:2018miv,Radice:2018ozg,Kiuchi:2019lls}, while they are also still in debate. To address these questions for GW170817 and also for the observations of the future events, quantitative understanding of the relation between the mass ejection mechanism and the EM counterparts is important. In this article, among the various types of EM counterparts for the NS mergers, we focus particularly on the EM counterparts that are launched by the consequence of sub- and mildly- relativistic mass ejection associated with the presence of a long-lived MNS.

Various studies in terms of the numerical simulations have been performed for NS mergers, revealing the detailed property of the ejecta and the resulting element abundances of the nucleosynthesis together with the dependence on the mass ejection mechanism and the binary parameters, such as the NS mass and NS EOS~(\citealt{Hotokezaka:2012ze,Bauswein:2013yna,Wanajo:2014wha,Sekiguchi:2015dma,Foucart:2015gaa,Sekiguchi:2016bjd,Radice:2016dwd,Dietrich:2016hky,Bovard:2017mvn,Kiuchi:2017zzg,Dessart:2008zd,Metzger:2014ila,Perego:2014fma,Just:2014fka,Wu:2016pnw,Siegel:2017nub,Shibata:2017xdx,Lippuner:2017bfm,Fujibayashi:2017puw,Siegel:2017jug,Ruiz:2018wah,Fernandez:2018kax,Christie:2019lim,Perego:2019adq,Miller:2019dpt,Fujibayashi:2020qda,Fujibayashi:2020jfr,Fujibayashi:2020dvr,Bernuzzi:2020txg,Ciolfi:2020wfx,Vsevolod:2020pak,Foucart:2020qjb,Fernandez:2020oow,Mosta:2020hlh,Shibata:2021bbj,Shibata:2021xmo}; see~\citealt{Shibata:2019wef} for a review). Based on or motivated by the knowledge of the ejecta profile and the element abundances obtained by those simulations, radiative transfer simulations with the realistic heating rate and/or the detailed opacity calculations have been performed to predict the kilonova light curves in the last decade~\citep[e.g.,][]{Kasen:2013xka,Kasen:2014toa,Barnes:2016umi,Wollaeger:2017ahm,Tanaka:2017lxb,Wu:2018mvg,Kawaguchi:2018ptg,Hotokezaka:2019uwo,Kawaguchi:2019nju,Korobkin:2020spe,Bulla:2020jjr,Zhu:2020eyk,Barnes:2020nfi,Nativi:2020moj,Kawaguchi:2020vbf,Wu:2021ibi,Just:2021vzy,Curtis:2021guz}.

Here it is to be emphasized that for accurately deriving the light curve and spectrum of kilonovae, long-term hydrodynamics evolution with the time scale of $\gg10\,{\rm s}$ is crucial. At the time of  the ejecta formation ($\lesssim 10\,{\rm s}$), the ejected matter still has non-negligible amount of internal energy compared to its kinetic energy, and thus, the subsequent ejecta trajectory can be modified by the thermal pressure~\citep{Kastaun:2014fna}. This hydrodynamical effect in the subsequent evolution could be reflected in the light curve of the EM counterparts. Currently, there are only a limited number of studies that consider the long-term hydrodynamics evolution of ejecta to predict the property of the EM counterparts~\citep{Rosswog:2013kqa,Grossman:2013lqa,Fernandez:2014bra,Fernandez:2016sbf,Foucart:2021ikp,Kawaguchi:2020vbf,Wu:2021ibi}.

Recently, \cite{Shibata:2021xmo} performed a long-term simulation for binary-neutron-star (BNS) merger remnants using a general relativistic radiation magneto-hydrodynamics (GRRMHD) simulation code with a mean field dynamo term~\citep{Shibata:2021bbj}. They paid attention to the models in which the remnant MNS survives for a long period after the merger with $\gtrsim 3$\,s. For these systems, it is indicated that intrinsic magneto-hydrodynamics (MHD) effects such as the magneto-centrifugal effect~\citep{BP82} and magnetic-tower effect may play an important role for the mass ejection if the magnetic-field strength in the merger remnant is amplified and the high field strength is preserved for a time scale longer than the mass ejection time scale. For example, the magnetic field lines anchored at the remnant MNS induce the magneto-centrifugal effect~\citep{BP82} to the surrounding matter, which receives angular momentum from the MNS and a part of which becomes a fast outflow. In this paper, we employ the numerical results of \cite{Shibata:2021xmo} to predict a variety of the hypothetical EM counterparts. We find that, in the presence of the significant MHD effects, a synchrotron afterglow with high luminosity can be associated with the merger event. We also perform a radiative transfer simulation for kilonovae and find that they can also be bright in the early epoch, while it shows the optical emission rapidly declining in a few days and the long-lasting ($\sim 10\,{\rm d}$) emission very bright in the NIR wavelength.

This paper is organized as follows: In Section~\ref{sec:method}, we describe the method employed in this study. In Section~\ref{sec:model}, we describe the BNS models we study in this work. The results of the nucleosynthesis calculation are also presented in this section. In Section~\ref{sec:result}, we present the property of the ejecta obtained by the long-term hydrodynamics evolution and the numerical results for the light curves of the EM counterparts. Finally, we discuss
the implication of this paper in Section~\ref{sec:dis}. Throughout this paper, $c$ denotes the speed of light.

\section{Method}\label{sec:method}
To obtain the ejecta profile in the homologously expanding phase, we evolve the ejecta for $\approx 0.1$\,d by solving the axisymmetric hydrodynamics equations employing the outflow data obtained by numerical-relativity (NR) simulations as the boundary condition~\citep{Shibata:2021xmo}. In the following, to distinguish between the present simulation and NR simulation, we refer to the present hydrodynamics simulations as the HD simulations.

We basically use the same method as that described in~\cite{Kawaguchi:2020vbf} for the HD simulations except for an update on the treatment of the radioactive heating effect. In this method, relativistic hydrodynamics equations in the spherical coordinates are solved taking into account the effect of fixed-background gravity of a non-rotating BH metric in the isotropic coordinates. We employ the ideal-gas EOS with the adiabatic index of $\Gamma=4/3$. In the previous work, the radioactive heating effect is incorporated by adding a source term in the energy and momentum equations. However, as noted in Appendix A of~\cite{Kawaguchi:2020vbf}, this naive treatment violates the energy and momentum conservation of the system: Because the radioactive heating is the consequence of the release of the binding energy of nuclei, the total energy and momentum of the system do not change by this process. Although the error due to the naive treatment is small for our setup, we have corrected our HD simulation code to self-consistently solve the system. The detail of the modification is summarized in Appendix~\ref{app:heat}. 

For the HD simulations, the uniform grid spacing with $N_\theta$ grid points is prepared for the polar angle $\theta$, while for the radial direction, the following non-uniform grid structure is employed; the $j$-th radial grid point is given by
\begin{align}
	{\rm ln}\,r_j={\rm ln}\left(\frac{r_{\rm out}}{r_{\rm in}}\right)\frac{j-1}{N_r}+{\rm ln}\,r_{\rm in},\,j=1\cdots N_r+1,\label{eq:grid}
\end{align} 
where $r_{\rm in}$ and $r_{\rm out}$ denote the inner and outer radii of the computational domain, respectively, and $N_r$ denotes the total number of the radial grid points. We set the same time origin for the HD simulations as in the NR simulations for the post-merger evolution. 

The time-sequential hydrodynamics property of the outflow is extracted from the NR simulations of~\cite{Shibata:2021xmo} at a certain radius, and is imposed at $r=r_{\rm in}$ of the HD simulations as the inner boundary condition. Accordingly, $r_{\rm in}$ is initially set to agree with the radius at which the outflow profile of the NR simulations is extracted. After the NR simulation data are run out at $t\sim 5$\,s, the HD simulation is continued by setting a very small floor-value to the density of the inner boundary. When the high velocity edge of the outflow reaches the outer boundary of our HD simulation, the radial grid points are added to the outside of the original outer boundary. At the same time, the innermost radial grid points are removed to keep the total number of the radial grid points. By this prescription, the value of $r_{\rm in}$ is increased in the late phase of the HD simulations, and $r_{\rm out}$ is always set to be $10^3 r_{\rm in}$. We note that the total mass in the removed grids lost by this prescription is always negligible ($\lesssim 4\times 10^{-4}\,M_\odot$) compared to the post-merger ejecta present around the inner boundary ($\sim 0.1\,M_\odot$).

For calculating the kilonova light curves and the synchrotron emission that arises from the ejecta fast tail, the ejecta profile at $0.1\,{\rm d}$ after the onset of the merger is employed. We perform the HD simulations with a larger grid resolution than the previous study, that is, $(N_r,N_\theta)=(3192,196)$ to resolve the fast tail of the ejecta with relatively low density and with mildly relativistic velocity. To predict the velocity distribution, which is employed to calculate the non-thermal emission, we switch off the radioactive heating effect in this HD simulation to reduce the computational time. 
This is justified by the fact that the energy deposition due to radioactive heating is negligible compared to the kinetic energy for the fast tail. Indeed, we confirmed that the light curves of the non-thermal emission shown below are not affected by the presence of the heating terms in the HD simulations (see Appendix~\ref{sec:conv}). 
For calculating kilonova light curves, we employ the same NR ejecta profile data as that for calculating the fast-tail, but the HD simulations are performed taking the radioactive heating into account, because the internal energy deposited by the radioactive heating during the ejecta expansion partially contributes to the kilonova light curves~\citep{Kawaguchi:2020vbf}. We note that a relatively small grid resolution of $(N_r,N_\theta)=(1024,128)$ is employed for the viscous model (see the next subsection), because we find that it is sufficient to resolve the ejecta with a small amount of the fast tail components ($v>0.6\,c$ with $v$ being the radial velocity).


\begin{figure*}
 	 \includegraphics[width=.5\linewidth]{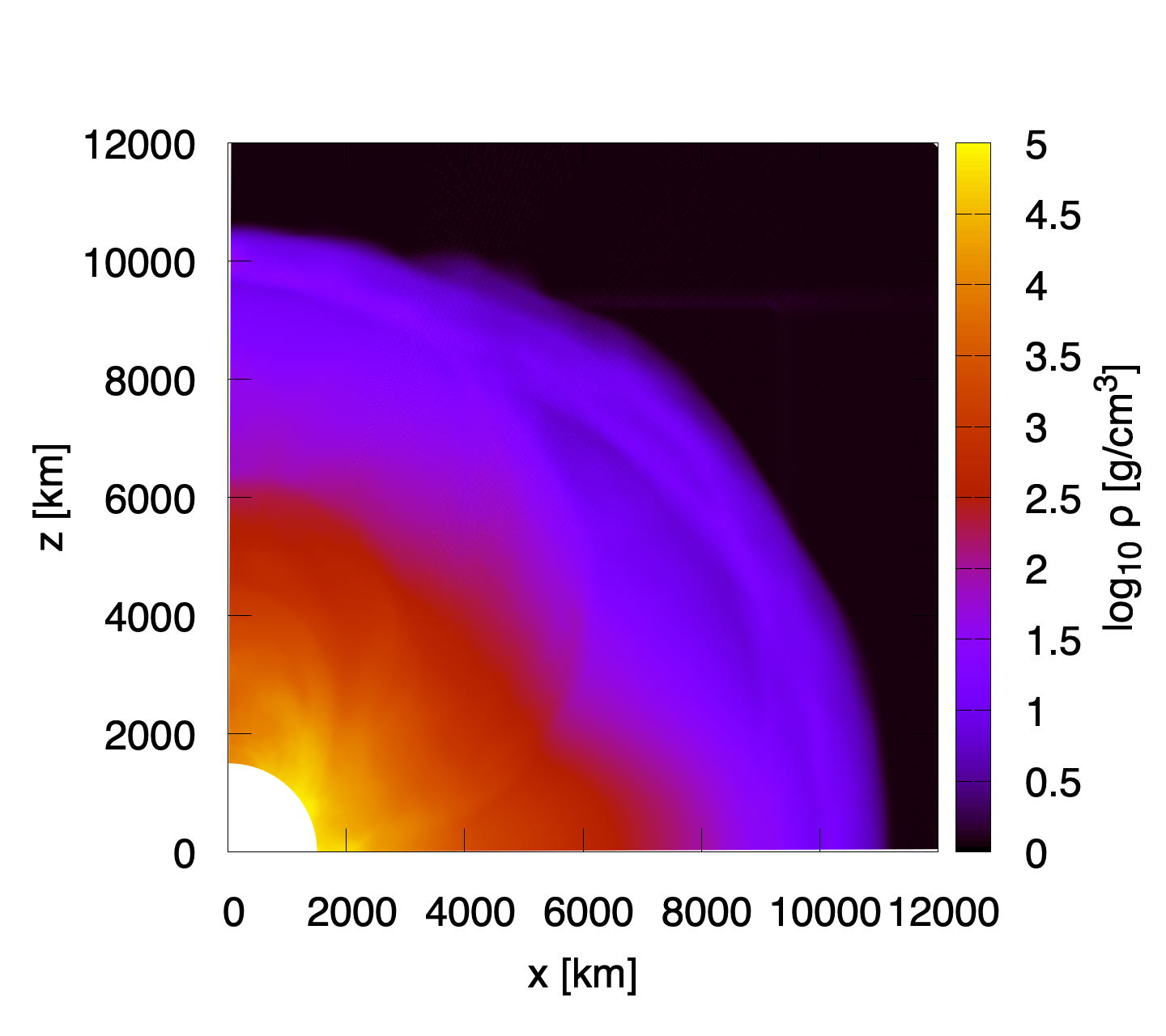}
 	 \includegraphics[width=.5\linewidth]{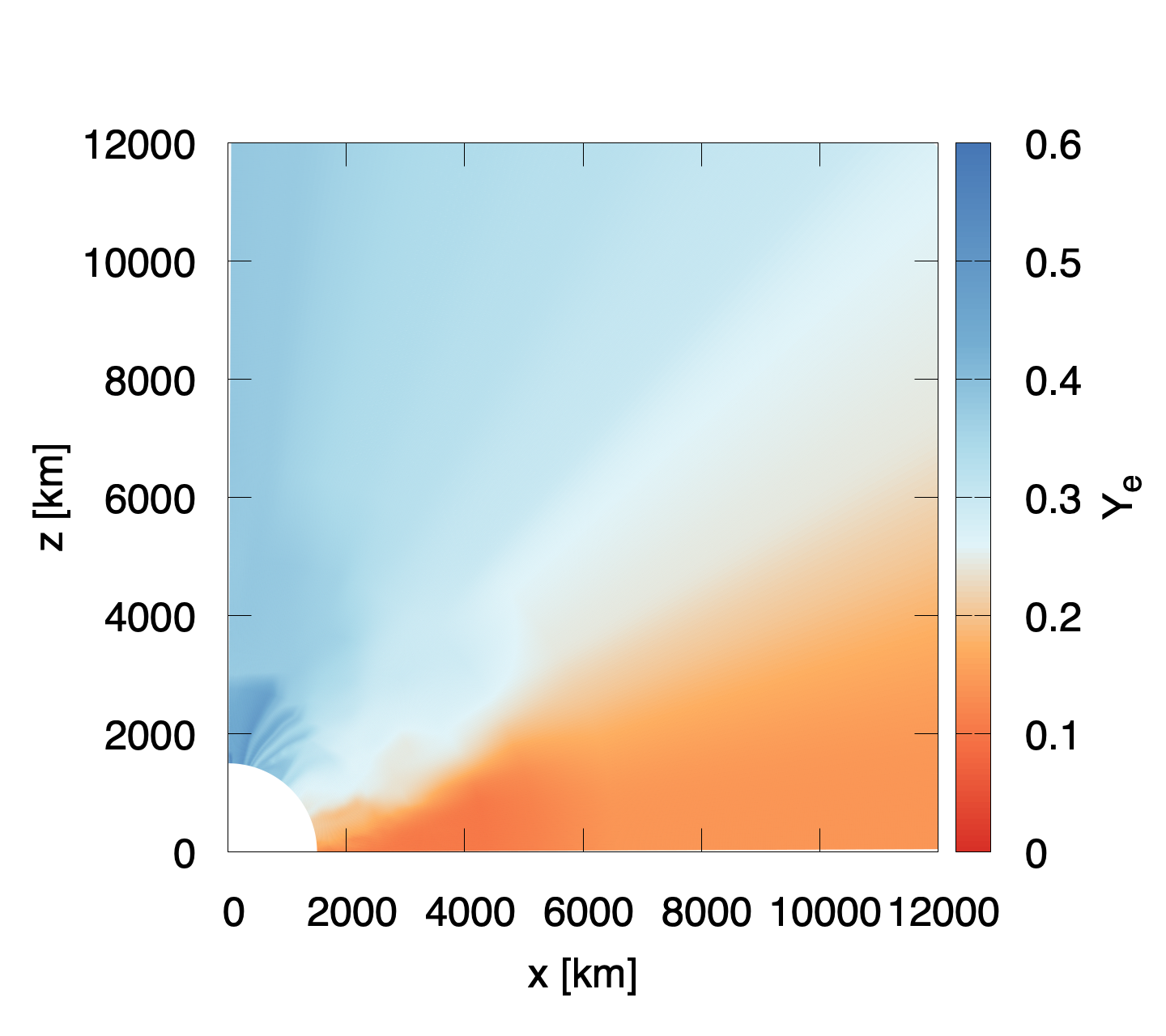}
 	 \caption{Profiles of ejecta rest-mass density (the left panel) and electron fraction (the right panel) for model DD2-135~\citep{Fujibayashi:2020dvr} of $r\geq1500\,{\rm km}$ at $t=0$. For the electron fraction profile, the value at the temperature of $5\times 10^9\,{\rm K}$ is shown. 
 	 }
	 \label{fig:ini}
\end{figure*}

\section{Model}\label{sec:model}
\subsection{Numerical relativity simulation}
\begin{table*}
\begin{center}
\caption{Key model parameters. The columns describe the model name, the phenomenological dynamo parameters of the NR simulations (see~\citealt{Shibata:2021xmo}), ejecta mass, ejecta energy, lanthanide+actinide mass fraction, and Y+Zr mass fraction, respectively. The values for the ejecta mass and energy shown outside and inside the parenthesis denote the values calculated from the input outflow data of the NR simulations and the results of the HD simulations at $t=0.1\,{\rm d}$, respectively. The values listed here are obtained by the results for the HD simulations without taking the radioactive heating into account.}
\begin{tabular}{c|c|c|c|c|c}\hline
Model	&	$(\sigma_{\rm c}\,[{\rm s^{-1}}],\alpha_{\rm d})$&	$M^0_{\rm eje}(M^{0.1\,{\rm }}_{\rm eje})\,[10^{-2}M_\odot]$&	$E_{\rm eje}(E^{0.1\,{\rm d}}_{\rm K,eje})\,[10^{51}M_\odot c^2]$&	$X_{\rm lan+act}$&	$X_{\rm Y+Zr}$\\\hline\hline
MNS70a	&	$(1\times10^{7},1\times10^{-4})$&	$4.6 (3.7)$&	$1.0 (0.84)$&	$6.9\times 10^{-3}$&	$7.6\times 10^{-2}$\\
MNS70b	&	$(1\times10^{7},2\times10^{-4})$&	$7.8 (6.1)$&	$2.1 (1.7)$&	$3.3\times 10^{-3}$&	$9.3\times 10^{-2}$\\
MNS75a	&	$(3\times10^{7},1\times10^{-4})$&	$9.4 (8.4)$&	$14 (13)$&	$4.2\times 10^{-3}$&	$1.2\times 10^{-1}$\\
MNS75b	&	$(3\times10^{7},2\times10^{-4})$&	$12 (12)$&		$16 (15)$&	$1.2\times 10^{-2}$&	$1.2\times 10^{-1}$\\
MNS80	&	$(1\times10^{8},1\times10^{-4})$&	$13 (13)$&		$41 (42)$&	$9.0\times 10^{-3}$&	$1.3\times 10^{-1}$\\
viscous ($\alpha=0.04$)&					---	&	$7.6(6.5)$&	$0.57(0.52)$&	$3.1\times 10^{-3}$&	$5.0\times 10^{-2}$\\\hline
\end{tabular}
\label{tb:model}
\end{center}
\end{table*}

As the input for the HD simulations, we employ the outflow profiles obtained by NR simulations for a post-merger evolution of a BNS in~\cite{Fujibayashi:2020dvr,Shibata:2021xmo}. The key quantities of each model are summarized in Table~\ref{tb:model}. The first five models listed in Table~\ref{tb:model} are obtained by long-term resistive MHD simulations with a mean-field dynamo term~\citep{Shibata:2021xmo}, employing the axisymmetrized merger remnant profile of an equal-mass BNS merger simulation with the DD2 EOS~\citep{Banik:2014qja} and with each NS mass of $1.35\,M_\odot$: see Table~\ref{tb:model} for the employed conductivity $\sigma_{\rm c}$ and mean-field dynamo parameters $\alpha_{\rm d}$. We note that, while the magnetic field energy reaches $\sim10^{51}\,{\rm erg}$ ($\sim1 \%$ of the kinetic energy of the system) at the peak for all the MHD models, for the larger values of $\sigma_{\rm c}$ the dissipation time scale of the magnetic fields is longer and for the larger values of $\alpha_{\rm d}$ the amplification time scale of the magnetic-field strength associated with the dynamo action is shorter. For the $\alpha_{\rm d}$ and $\sigma_{\rm c}$ values employed in the models listed in Table~\ref{tb:model}, the magnetic field amplification time scale and dissipation time scale are found to be $\sim10\,{\rm ms}$ and $\sim 1\,{\rm s}(\sigma_{\rm c}/10^8\,{\rm s}^{-1})$, respectively (see Section IIB and Figure 9 in~\citealt{Shibata:2021xmo}). The last model listed in Table~\ref{tb:model} with the label of ``$\alpha =0.04$" denotes the result for the same BNS model but of a viscous-hydrodynamics simulation with the dimensionless alpha viscous parameter of $\alpha=0.04$~\citep{Fujibayashi:2020dvr}.

Table~\ref{tb:model} summarizes the ejecta mass, $M^0_{\rm eje}$, and ejecta energy, $E^0_{\rm eje}$, calculated from the time sequence of the input outflow data of the NR simulations by the following equations:
\begin{align}
	M^0_{\rm eje}&:=\int_{r=r_{\rm in}} \rho_*v^r dt dS,\\
	E^0_{\rm eje}&:=\int_{r=r_{\rm in}} \rho_* c^2 (hw-1)v^r dt dS.
\end{align}
Here, $\rho_*:=\rho u^t \sqrt{-g}$ with $\rho$ the rest-mass density, $u^t$ the time component of the four velocity $u^\mu$, and $g$ the determinant of the spacetime metric. $v^r:=u^r/u^t$, $w:=\alpha_g u^t$ with $\alpha_g$ the lapse function, and $h$ is the specific enthalpy\footnote{More precisely, $\rho_*$ and $h$ correspond to ${\hat \rho}_*$ and ${\hat h}$ in Appendix~\ref{app:heat}, respectively. In the following, we omit the hat symbol ``\textasciicircum" for these variables for the readability.}. We note that $E^0_{\rm eje}$ contains the contribution from both kinetic- and internal-energy flows. We note that all the matter crossing the inner radius is included in the integration. The contribution from the dynamical ejecta to the total ejecta mass and ejecta energy, {which are determined only by the merger phase, are $\approx 1.5\times10^{-3}\,M_\odot$ and $\approx 5\times10^{49}\,{\rm erg}$, respectively~\citep{Fujibayashi:2020dvr}. Thus, the post-merger ejecta contributes primarily to these quantities. 

As found from Table~\ref{tb:model}, the ejecta become massive and more energetic as the value of $\alpha_{\rm d}$ or $\sigma_{\rm c}$ increases (i.e., as the post-merger MHD activity is enhanced). In particular, the increase in the ejecta energy is significant as the value of $\alpha_{\rm d}$ increases, while difference in the ejecta mass is within a factor of 3 among the models. Since the rotational kinetic energy of the MNS ($\sim10^{52-53}\,{\rm erg}$) is the main source to accelerate the ejecta through the magneto-centrifugal effect~\citep{BP82} in the MHD simulations, these results indicate that the efficiency of converting the rotational kinetic energy of the MNS to the kinetic energy of the ejecta depends strongly on the magnetic-field evolution. Specifically, the dissipation time scale of the magnetic fields controlled by the conductivity determines the efficiency. On the other hand, the ejecta mass converges within $20$--$50$\% of the disk mass irrespective of the dynamo parameters~\citep{Shibata:2021xmo}.
The ejecta mass for the viscous model (with $\alpha=0.04$) is as large as for model MNS70b and larger than for model MNS70a. However the ejecta energy for the viscous model has only a half of that for MNS70a, for which the MHD activity is weakest among the present models. This also shows that the kinetic energy of the ejecta is enhanced by an intrinsic MHD effect. 

Figure~\ref{fig:ini} shows the profiles of the rest-mass density and electron fraction at $t=0$ of the HD simulations (note that the HD and NR simulations for the post-merger evolution share the same time origin). We note that, for the electron fraction profile, the value at the temperature of $5\times 10^9\,{\rm K}$ is shown. In this initial time slice, only the dynamical ejecta component is present in the computational domain. The rest-mass density profile of the dynamical ejecta exhibits weakly spheroidal morphology, although the electron fraction is distributed in an anisotropic manner. The ejecta in the polar region of $\theta\alt45^\circ$--$60^\circ$ are dominated by the matter with $Y_{\rm e}\geq0.3$, while those around the equatorial region ($\theta\agt60^\circ$) are dominated with $Y_{\rm e}\leq0.2$. The difference in the ejecta electron fraction reflects the difference in the mass ejection mechanisms: the polar component is driven by the collisional shock heating during the merger of two NSs, while the other component is driven by the tidal torque induced by the non-axisymmetric matter distribution at the onset of the merger~\citep{Wanajo:2014wha,Sekiguchi:2015dma,Sekiguchi:2016bjd,Radice:2016dwd}.

\subsection{Nucleosynthesis}


\begin{figure*}
 	 \includegraphics[width=.5\linewidth]{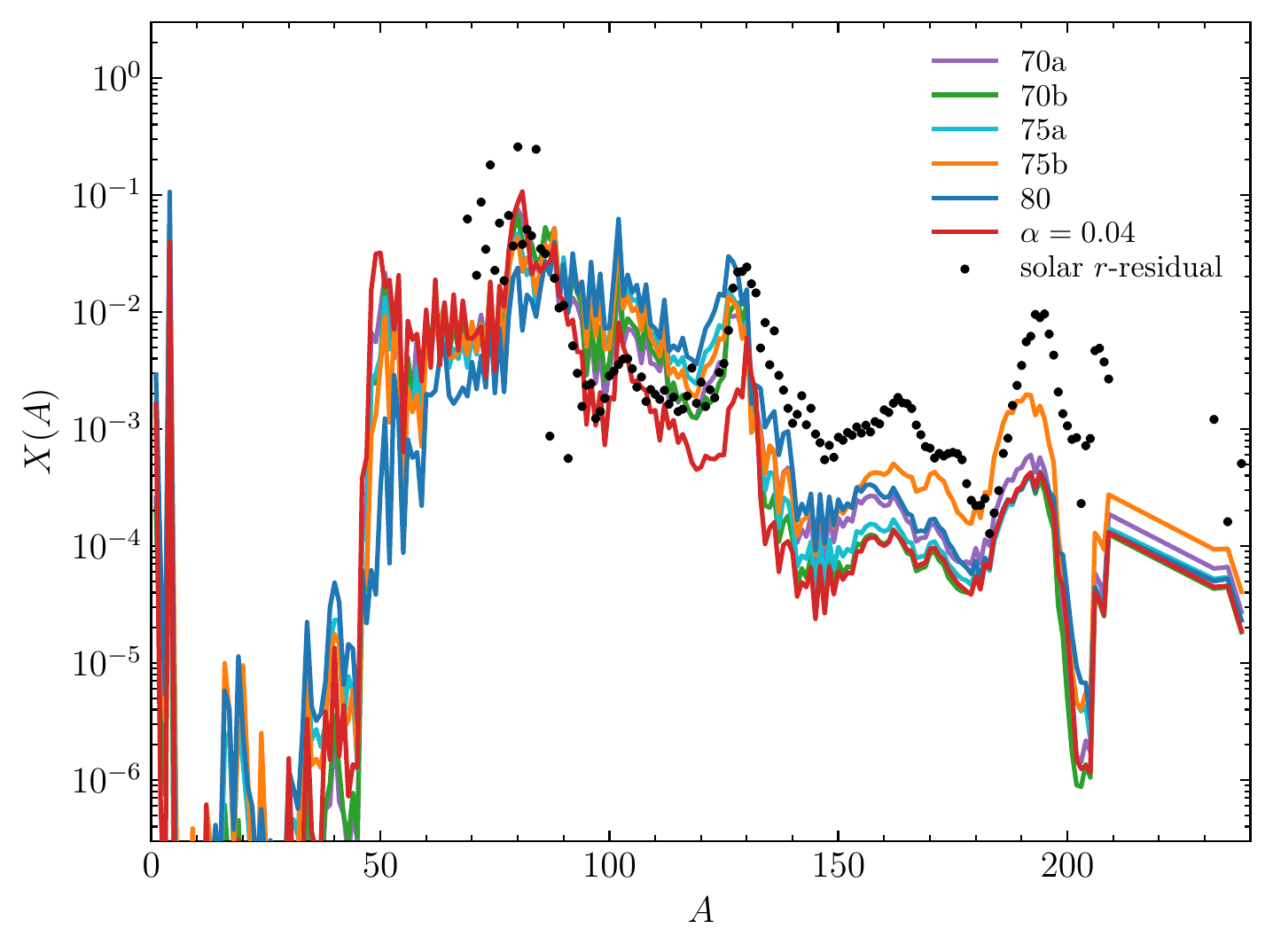}
 	 \includegraphics[width=.5\linewidth]{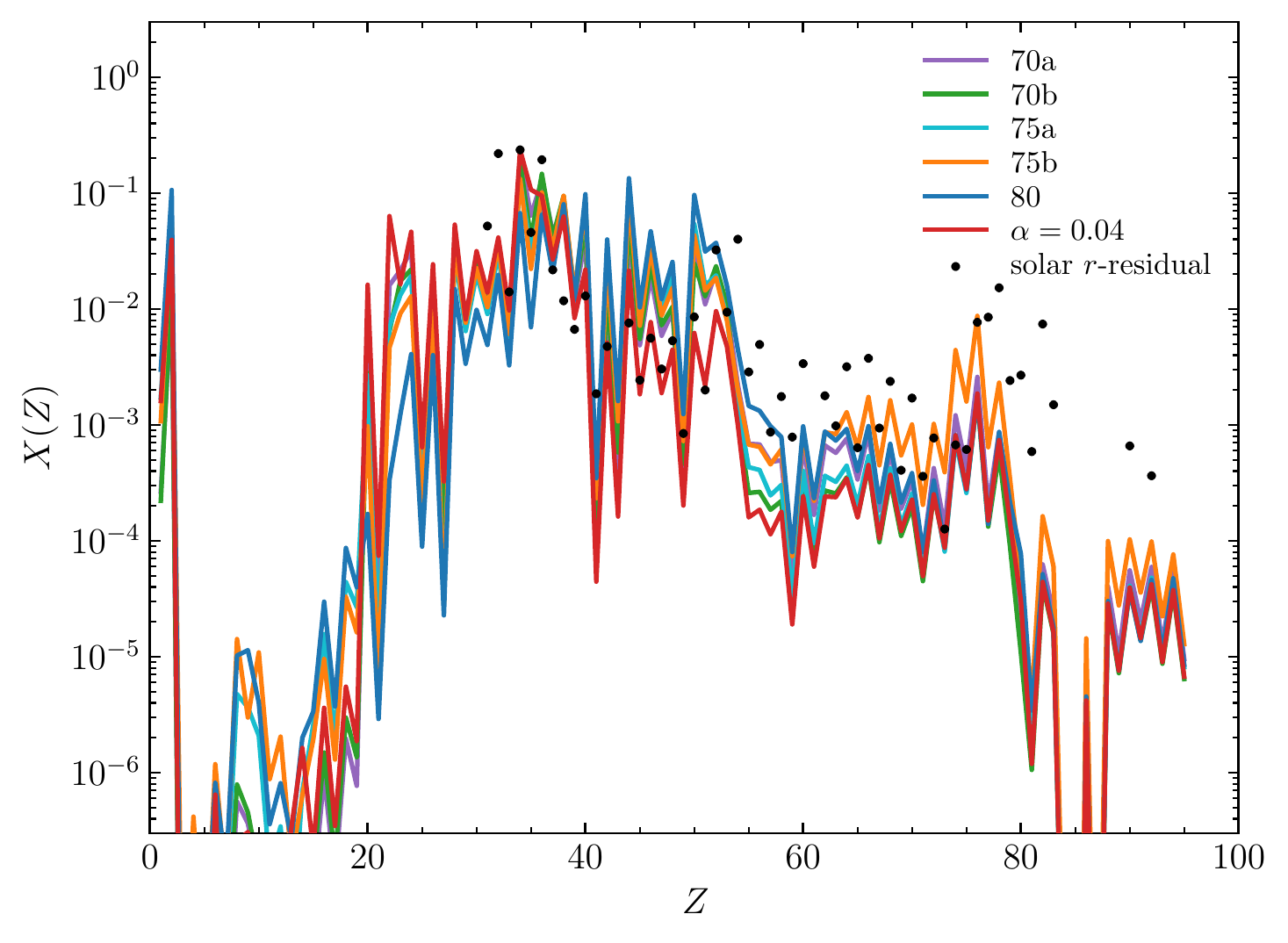}
 	 \caption{Mass fractions as a function of atomic mass number $A$ (left; after decay) and of atomic number $Z$ (right; at 1 d) for all the models. The solar $r$-residuals (denoted by black circles) are vertically shifted to match that for the case of $\alpha = 0.04$ at $A = 82$ and $Z = 34$ in the left and right panels, respectively.}
	 \label{fig:nucleosynthesis}
\end{figure*}

We performed a nucleosynthesis calculation in each ejecta particle for a given model. The method of the nucleosynthesis calculation as well as the post-process particle tracing is the same as those described in our latest paper \citep{Fujibayashi:2020dvr}. In this work, we set the radius for the extraction of the ejecta information at 1500\,km that is the initial value of $r_\mathrm{in}$ for the HD simulations. The tracer particles are distributed at this radius ($r=1500$\,km) uniformly in the polar angle, which are traced back in time. The bulk of the dynamical ejecta has already passed the extraction radius at the beginning of the NR (MHD) simulations (see Fig.~\ref{fig:ini}). We adopt, therefore, the result of the nucleosynthesis calculation in the NR (viscous) model of \citet{Fujibayashi:2020dvr} for the matter with initial radius $1500\,{\rm km}<r<8000\,{\rm km}$. 



Figure~\ref{fig:nucleosynthesis} shows the results of the nucleosynthesis calculation for all the models listed in Table \ref{tb:model}. The left and right panels present the mass fractions as functions of atomic mass number $A$ (after decay) and of atomic number $Z$ (at 1 d), respectively, the latter being used for the radiative transfer simulations in this study. The heavy $r$-process elements are under-produced than the solar abundance pattern ($r$-process residuals to the solar abundance taken from \citealt{Prantzos:2020}) not only for the viscous-hydrodynamics model but also for the MHD models irrespective of the dynamo parameters. This is because the mass of the post-merger ejecta is by more than one order of magnitude larger than that of the dynamical ejecta and the post-merger ejecta predominantly synthesizes relatively light $r$-process elements with the atomic mass number smaller than 130. On the other hand, significant amounts of Y ($Z=39$) and Zr ($Z=40$), which contribute a lot to the opacity in the optical wavelength ~\citep{Tanaka:2019iqp,Kawaguchi:2020vbf,Ristic:2021ksz}, are synthesized (see also Table~\ref{tb:model} for the lanthanide+actinide and Y+Zr mass fraction of the ejecta).
Note that the MHD models synthesize a slightly larger amount of $r$-process elements (including lanthanides+actinides and Y+Zr) than that in the viscous model ($\alpha = 0.04$) owing to the higher entropies in the ejecta (and in part higher outgoing velocities) for the formers \citep{Shibata:2021xmo}.

As we discussed in \cite{Fujibayashi:2020dvr}, the abundance pattern shown in Figure~\ref{fig:nucleosynthesis} is universally found in the presence of long-lived MNSs as a merger remnant. This abundance pattern does not agree with the solar abundance pattern because of the overproduction of the relatively light elements with $A \alt 130$ (underproduction of the heavy elements). Therefore, suppose that the solar abundance pattern is universal in the universe and the main site for the $r$-process nucleosynthesis is BNS mergers, the channel in which a long-lived MNS is formed should be the minority of the BNS mergers (but see, Refs.~\citealt{Wu:2017qpc,Wu:2017drk,Li:2021vqj}, for the case that the neutrino oscillation plays an important role in determining the ejecta $Y_\mathrm{e}$). In other words, if the future observation suggests that the formation of the long-lived remnant MNSs is the majority of the BNS mergers, it will indicate that BNS mergers might not be the main production sites of the $r$-process elements. Besides the event rates, however, this merger channel synthesizes rich $r$-process elements, and thus, is the promising source of bright kilonovae, which may be observed even if they appear at a large distance. We show the light-curve models of kilonovae in Section~\ref{sec4.3}.

\section{Results}\label{sec:result}

\begin{figure*}
 	 \includegraphics[width=.5\linewidth]{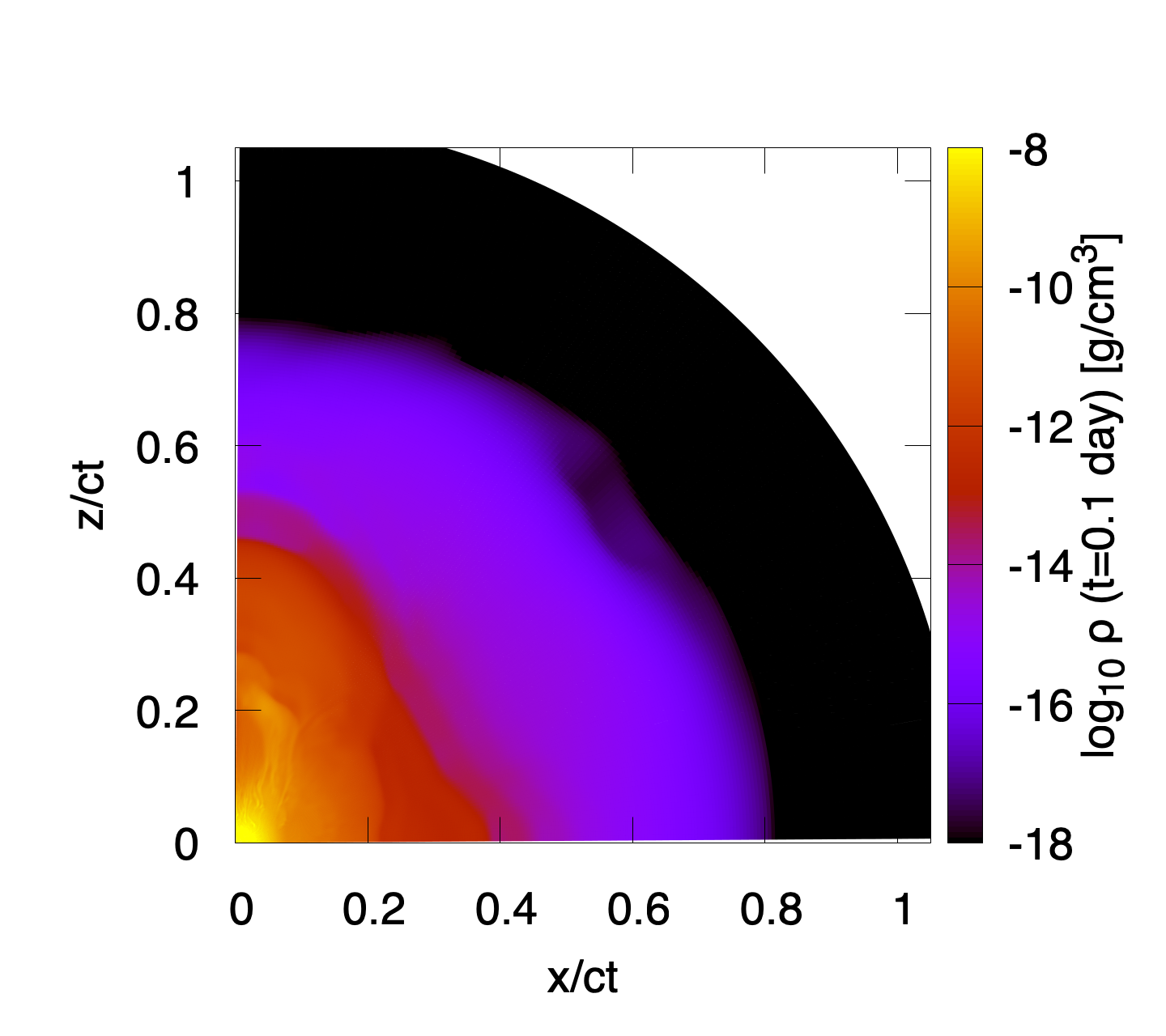}
 	 \includegraphics[width=.5\linewidth]{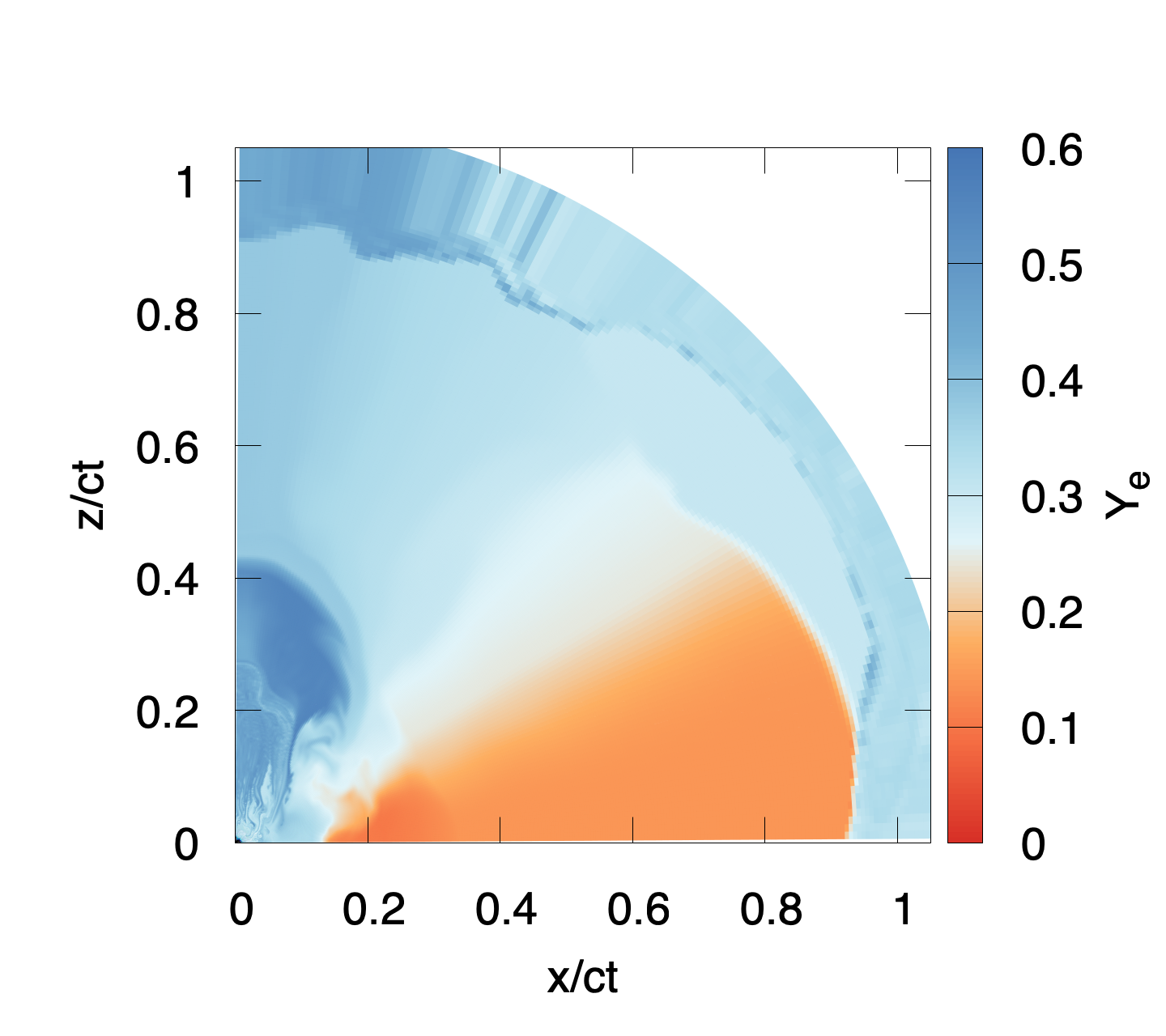}
 	 \caption{Profiles of ejecta rest-mass density (the left panel) and electron fraction (the right panel) at $t=0.1\,{\rm d}$ for the viscous model ($\alpha=0.04$). The value at the temperature of $5\times 10^9\,{\rm K}$ is shown for the electron fraction profile.}
	 \label{fig:profvis}
\end{figure*}

\subsection{Ejecta profiles}
This subsection focuses on the results of the ejecta profile in the HD simulation. As we found in the previous study, a fraction of matter which passed through the extraction radius in the NR simulations fails to become gravitationally unbound during the subsequent hydrodynamics evolution even if the often-used Bernoulli criteria ($\equiv hu_t\leq-h_{\rm min}$, see~\citealt{Kastaun:2014fna,Vincent:2019kor,Fujibayashi:2020dvr,Foucart:2021ikp}) was satisfied at the time of passing through. The reason for this is that the pressure of the precedingly outgoing matter can decelerate the matter that catches up. Hence, the total rest mass and energy of the ejecta at the homologously expanding phase can be slightly different from those measured in the NR simulations.\footnote{
Because the specific enthalpy contains a contribution of nuclear binding energy, we assume in this work that the released rest-mass energy by nuclear burning is totally used to accelerate matter in the Bernoulli criteria, setting that the asymptotic specific enthalpy is the minimum of the EOS table adopted (i.e., assuming that irons are produced). This treatment could overestimate the total mass of the unbound matter, because the released energy should be smaller if the elements other than irons (e.g., $r$-process elements) are produced.  
Moreover, only a part of the rest-mass energy released by the nuclear burning can be used to accelerate the matter, because a fraction of the energy is carried away by neutrinos: For the $r$-process nucleosynthesis, neutrinos carry away $\sim 40$\% of the energy \citep{Hotokezaka:2016,Foucart:2021ikp}. 
Due to these reasons, a fraction of the matter can still be asymptotically bound even if the Bernoulli criteria is satisfied.}

To quantify the rest mass and kinetic energy of the ejecta at the homologously expanding phase, we evaluate the following quantities at $t=0.1\,{\rm d}$: 
\begin{align}
	M^{0.1\,{\rm d}}_{\rm eje}&:=\left.\int \rho_*dV\right|_{t=0.1\,{\rm d}}\\
	E^{0.1\,{\rm d}}_{\rm K,eje}&:=\left.\int \rho_* c^2 (w-1)dV\right|_{t=0.1\,{\rm d}}.
\end{align}
Here, we simply measure the mass and the energy in the computational domain of the HD simulation, because, at $t=0.1\,{\rm d}$, approximately all the fluid elements satisfy the geodesic criteria of unbound matter ($\equiv u_t < -1$). The numerical results are listed in Table~\ref{tb:model}, which are obtained without taking the radioactive heating into account in the HD simulations. We note that the ejecta mass increases by $10$--$15\%$ if we take the radioactive heating into account, although it has only a negligible contribution to the EM light curves.

The differences between $M^0_{\rm eje}$ and $M^{0.1\,{\rm d}}_{\rm eje}$ and between $E^0_{\rm eje}$ and $E^{0.1\,{\rm d}}_{\rm K,eje}$ denote the rest mass and energy of the fall-back matter. Note that all the matter passing through the inner boundary (including that finally falls back) is counted to $M^0_{\rm eje}$  and $E^0_{\rm eje}$ and satisfies the Bernoulli criteria ($\equiv hu_t\leq-h_{\rm min}$, see~\citealt{Fujibayashi:2020dvr}) at the time of passing through. These quantities decrease as the values of $\alpha_{\rm d}$ and $\sigma_{\rm c}$ increase, i.e., the kinetic energy (and velocity) of the ejecta becomes higher. The fact that $E^{0.1\,{\rm d}}_{\rm K,eje}/E^0_{\rm eje}$ is larger than $M^{0.1\,{\rm d}}_{\rm eje}/M^0_{\rm eje}$ implies that the specific energy of the fall-back matter is smaller than the bulk specific energy of the ejecta.

We note that in the HD simulations, after the outflow data of the NR simulations run out, the rest-mass density on the inner boundary is switched to that for an atmosphere value ($=$ a sufficiently small floor value). We should note that this treatment could artificially increase the mass of the fall-back matter due to the sudden vanishing of the pressure support on the inner boundary. Nevertheless, as found in our previous paper~\cite{Kawaguchi:2020vbf}, the contribution of such marginally unbound matter to either the synchrotron afterglow or kilonova emission is minor because it has only low velocity.

\subsubsection{2D ejecta profile}

\begin{figure*}
 	 \includegraphics[width=.5\linewidth]{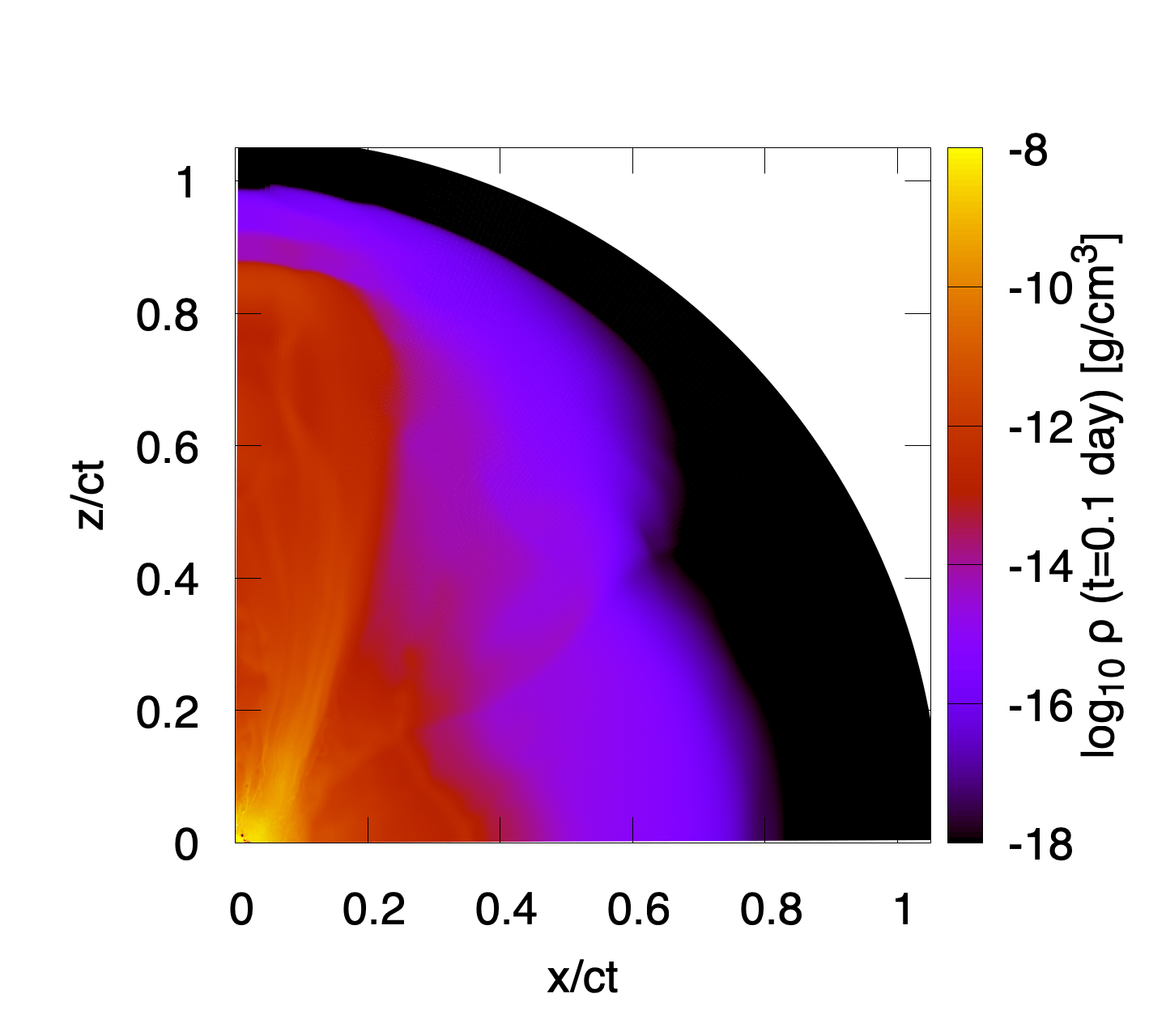}
 	 \includegraphics[width=.5\linewidth]{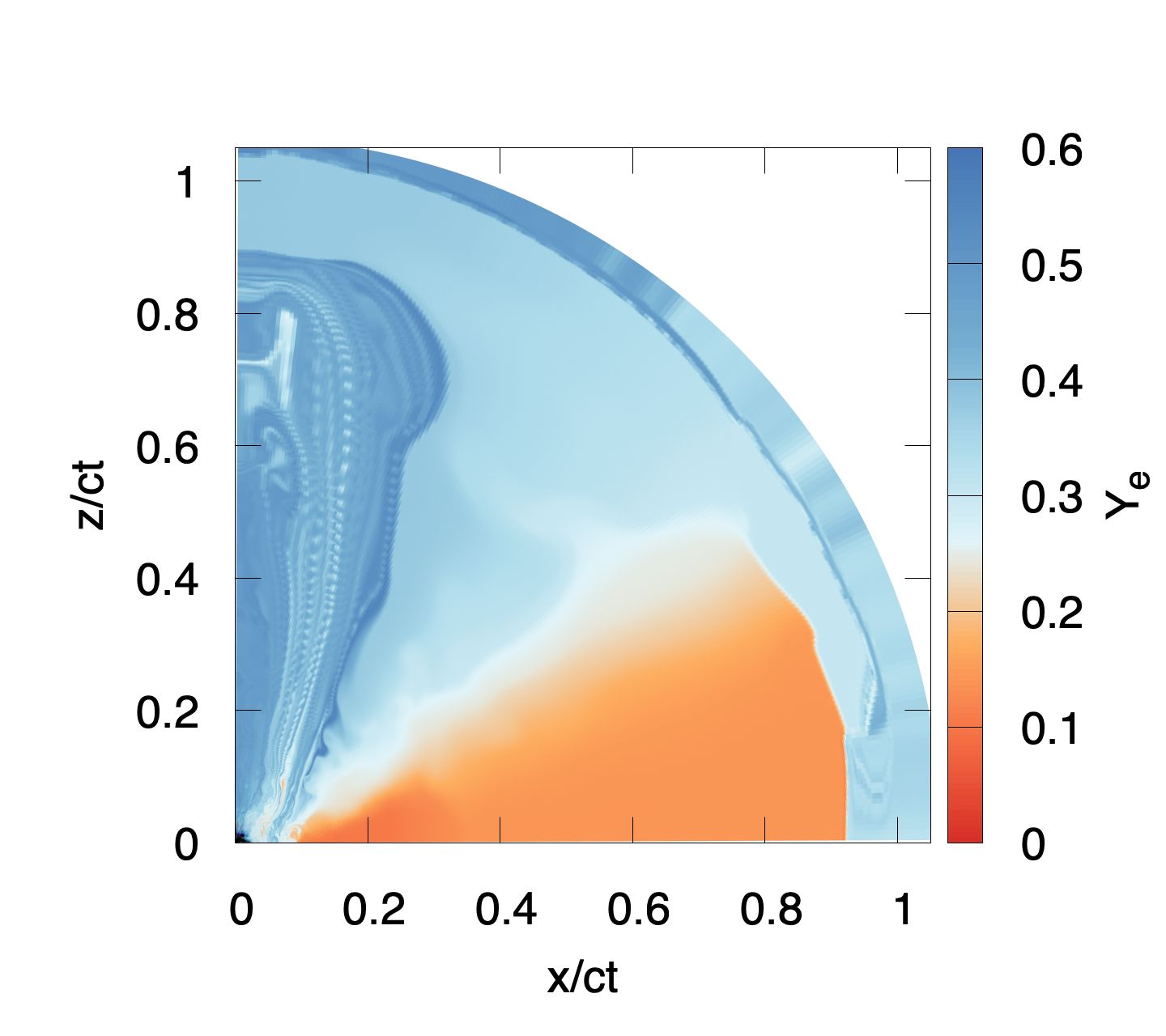}\\
 	 \vspace{-8mm}
 	 \includegraphics[width=.5\linewidth]{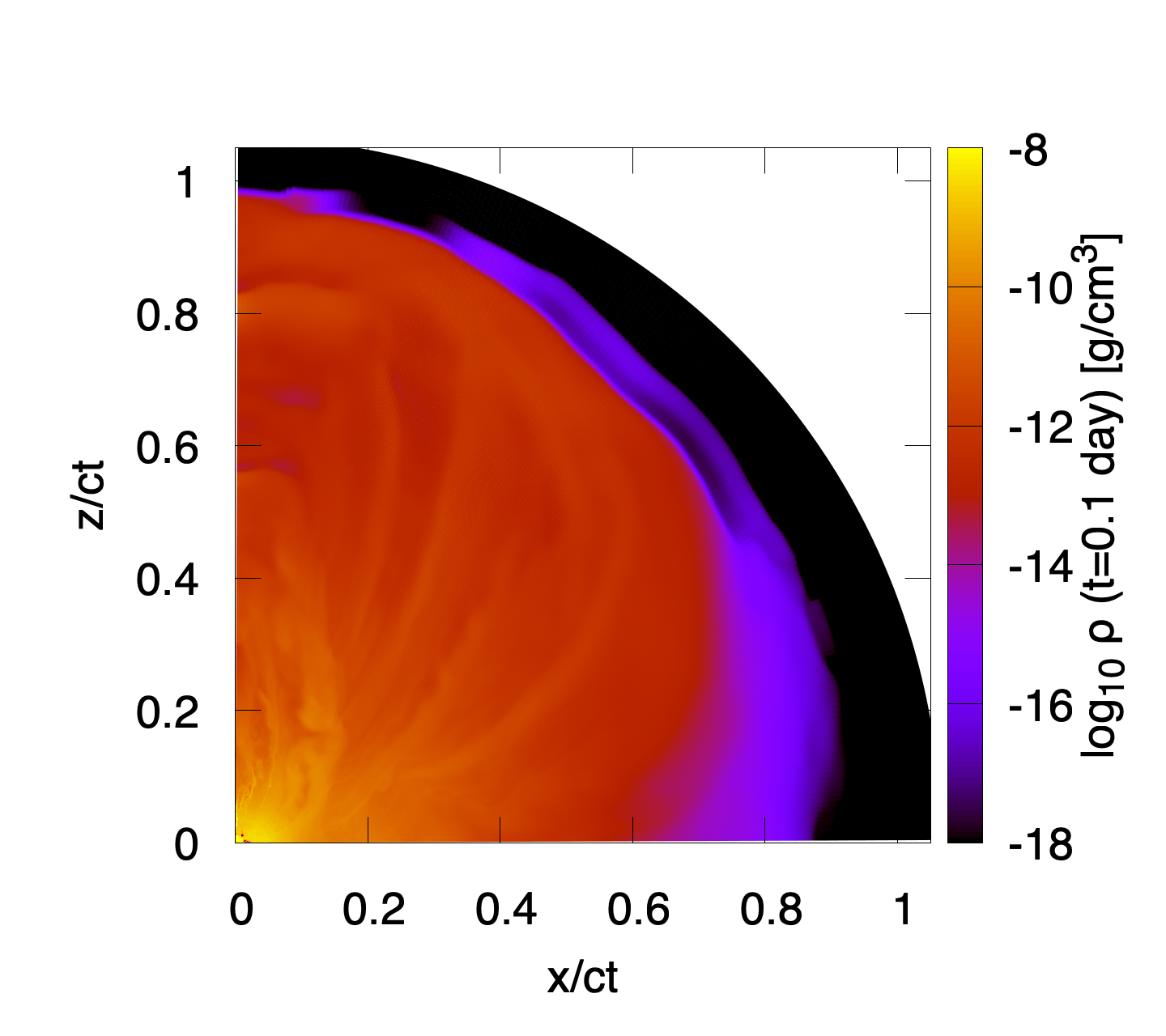}
 	 \includegraphics[width=.5\linewidth]{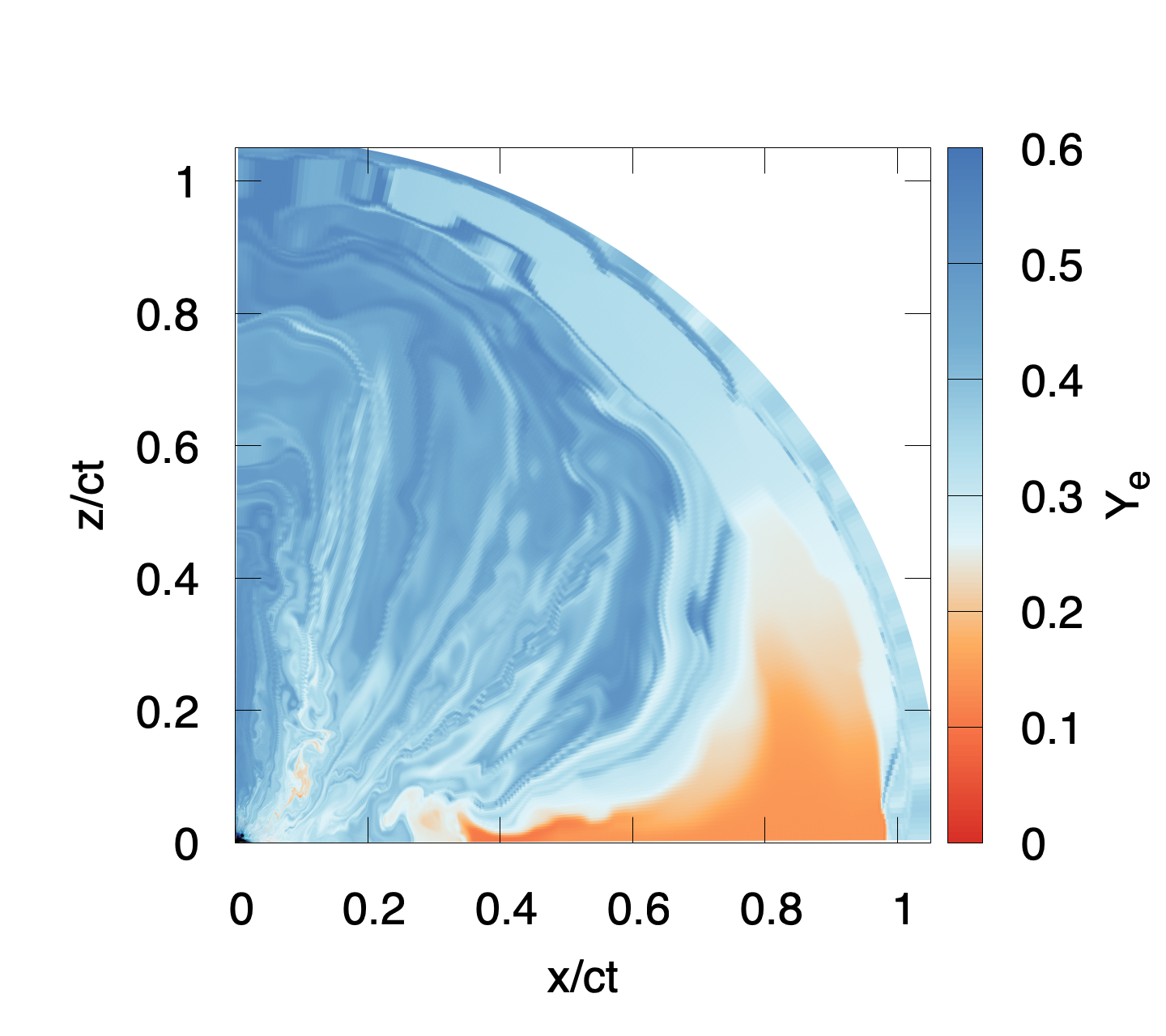}\\
 	 \vspace{-8mm}
 	 \includegraphics[width=.5\linewidth]{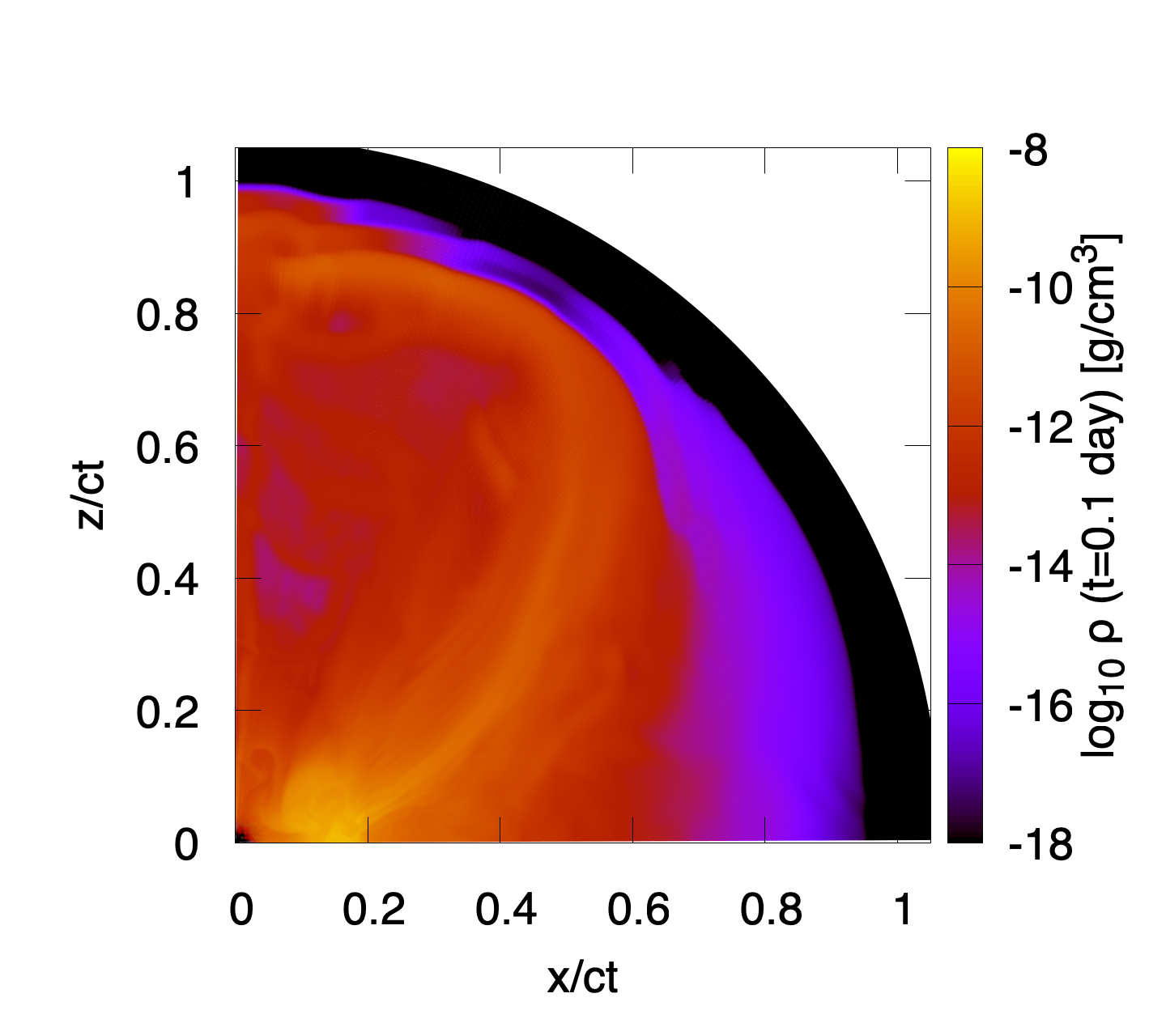}
 	 \includegraphics[width=.5\linewidth]{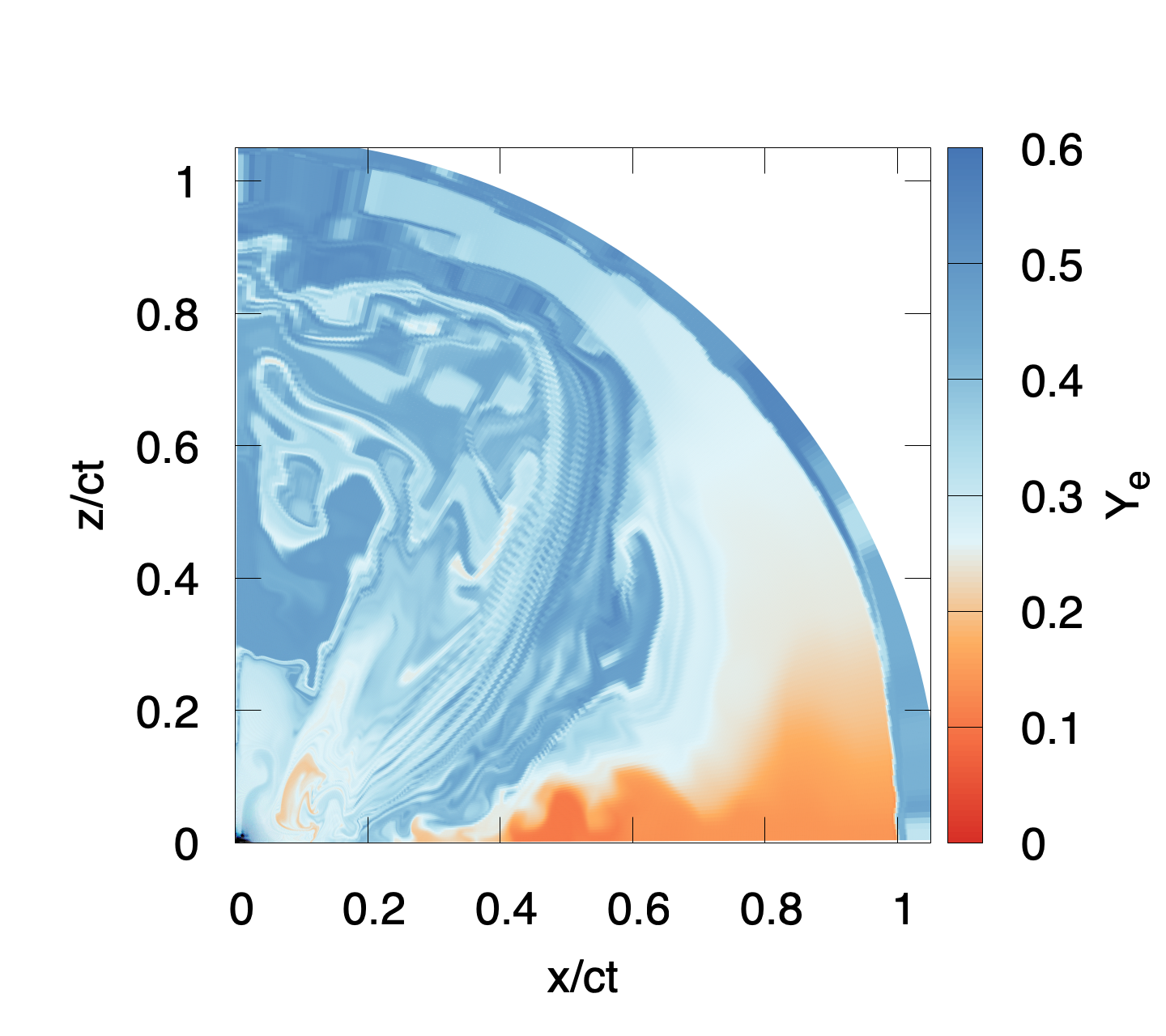}\\
 	 \caption{The same as Figure~\ref{fig:prof} but for the MHD models. The top, middle, and bottom panels denote the results for models MNS70a, MNS75a, and MNS80, respectively.}
	 \label{fig:prof}
\end{figure*}

Figures~\ref{fig:profvis} and~\ref{fig:prof} show the profiles of the rest-mass density and electron fraction of the ejecta at $t=0.1\,{\rm d}$ for the viscous model ($\alpha=0.04$) and for models MNS70a, MNS75a, and MNS80, respectively. We note that, as is the case for Figure~\ref{fig:ini}, the value at the temperature of $5\times 10^9\,{\rm K}$ is shown for the electron fraction profile. We note that the ejecta profiles shown in Figures~\ref{fig:profvis} and~\ref{fig:prof} are obtained in the HD simulations that take the radioactive heating into account. 

As \cite{Shibata:2021xmo} pointed out, the outflow property for the MHD models with a small value of the conductivity, $\sigma_{\rm c}=1\times 10^7\,{\rm s^{-1}}$, is qualitatively similar to that for the viscous model. Thus, we first discuss the results for the viscous model as the reference, and then, we compare the results for the MHD models with those for the viscous model to clarify the specific features of the MHD effects.

The rest-mass density and electron fraction profiles for the viscous model ($\alpha=0.04$) show approximately the same features as those found for the viscous model in the previous study (a low-mass model, DD2-125; \citealt{Fujibayashi:2020dvr,Kawaguchi:2020vbf}). 
The density profile of the ejecta with $r/ct \lesssim 0.1$ exhibits a weakly spheroidal shape. This component is driven by the viscous effect as the post-merger ejecta from the torus surrounding the central MNS. On the other hand, the density profile of the high-velocity ejecta with $r/ct \gtrsim 0.1$} exhibits mildly prolate morphology: For example, the rest-mass density larger than $10^{-13}\,{\rm g/cm^3}$ at $t=0.1\,{\rm d}$ is distributed for the velocity space up to $r/ct\approx0.5$ and $\approx0.4$ toward the polar and equatorial directions, respectively. For the $Y_{\rm e}$ distribution, primarily two distinct components are seen: One has a torus-like shape located around $x/ct\approx0.2$--$0.4$ and $\theta\agt45^\circ$ with $Y_{\rm e}\alt0.25$, and the other has a prolate shape located around $x/ct\alt0.2$ and $\theta\alt45^\circ$  with $Y_{\rm e}\agt0.3$. The former corresponds to the dynamical ejecta component driven by tidal torque induced by the non-axisymmetric matter distribution at the onset of the merger. The latter is composed of a part of the dynamical ejecta driven by the shock heating at the merger and neutrino-irradiation-enhanced ejecta~\citep{Fujibayashi:2020dvr,Kawaguchi:2020vbf}. 

In contrast to the viscous model, an appreciable amount of the ejecta with high velocity ($\agt0.5\,c$) is present in the MHD models, in particular for $\sigma_{\rm c}\geq 3 \times 10^7\,{\rm s}^{-1}$, for which the dissipation time of the magnetic field in the MNS is longer than or as long as the time scale of the post-merger mass ejection of several hundreds ms. Figure~\ref{fig:prof} shows that the high-velocity post-merger ejecta push the dynamical ejecta. The pushed ejecta component is highly appreciable for $\sigma_{\rm c}\geq 3 \times 10^7\,{\rm s}^{-1}$ and the highest velocity of the ejecta with the density larger than $10^{-13}\,{\rm g/cm^3}$ at $t=0.1\,{\rm d}$ reaches $\agt0.9\,c$ in the polar direction. The velocity of the post-merger ejecta can reach $\sim 0.7c$ even in the equatorial direction. As the consequence, the low-$Y_{\rm e}$ dynamical ejecta is accelerated and confined toward the equatorial plane, and also, the ejecta rest-mass density profile settles into a more isotropic shape than that for the viscous model. By contrast, the MHD model with a relatively small value of $\sigma_{\rm c}=1\times 10^7\,{\rm s}^{-1}$ (MNS70a) shows a ejecta profile  similar to those for the viscous model besides the presence of the high velocity ($\agt0.5\,c$) matter in the polar region. This similarity is due to the fact that for such a small value of $\sigma_{\rm c}$ the dissipation time scale of the magnetic fields in the MNS is shorter than the post-merger mass-ejection time scale and the acceleration of the post-merger ejecta by the magneto-centrifugal effect plays only a minor role.

\subsubsection{Kinetic energy distribution of ejecta}\label{sec4.1.2}
\begin{figure}
 	 \includegraphics[width=1.\linewidth]{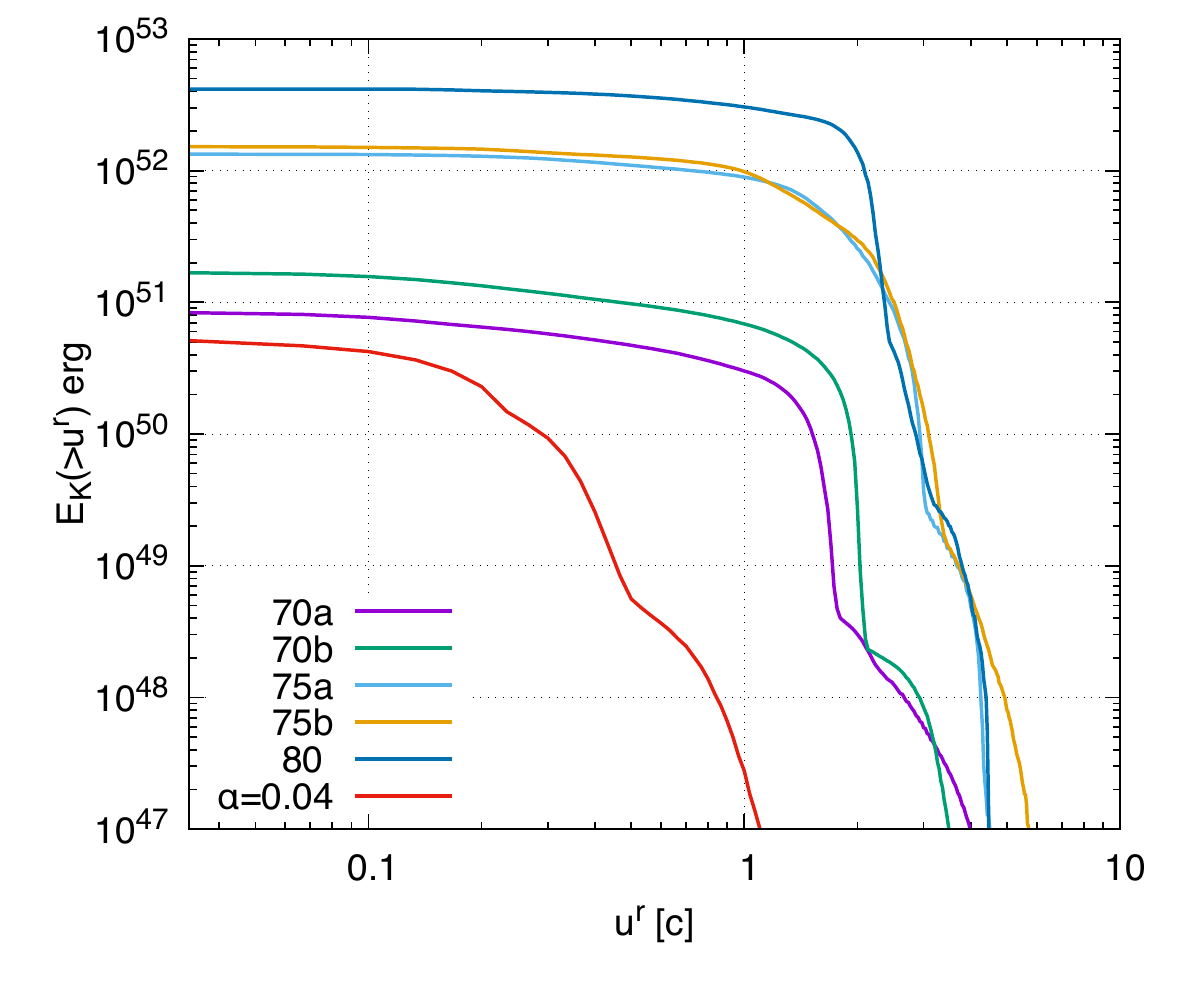}\\
 	 \includegraphics[width=1.\linewidth]{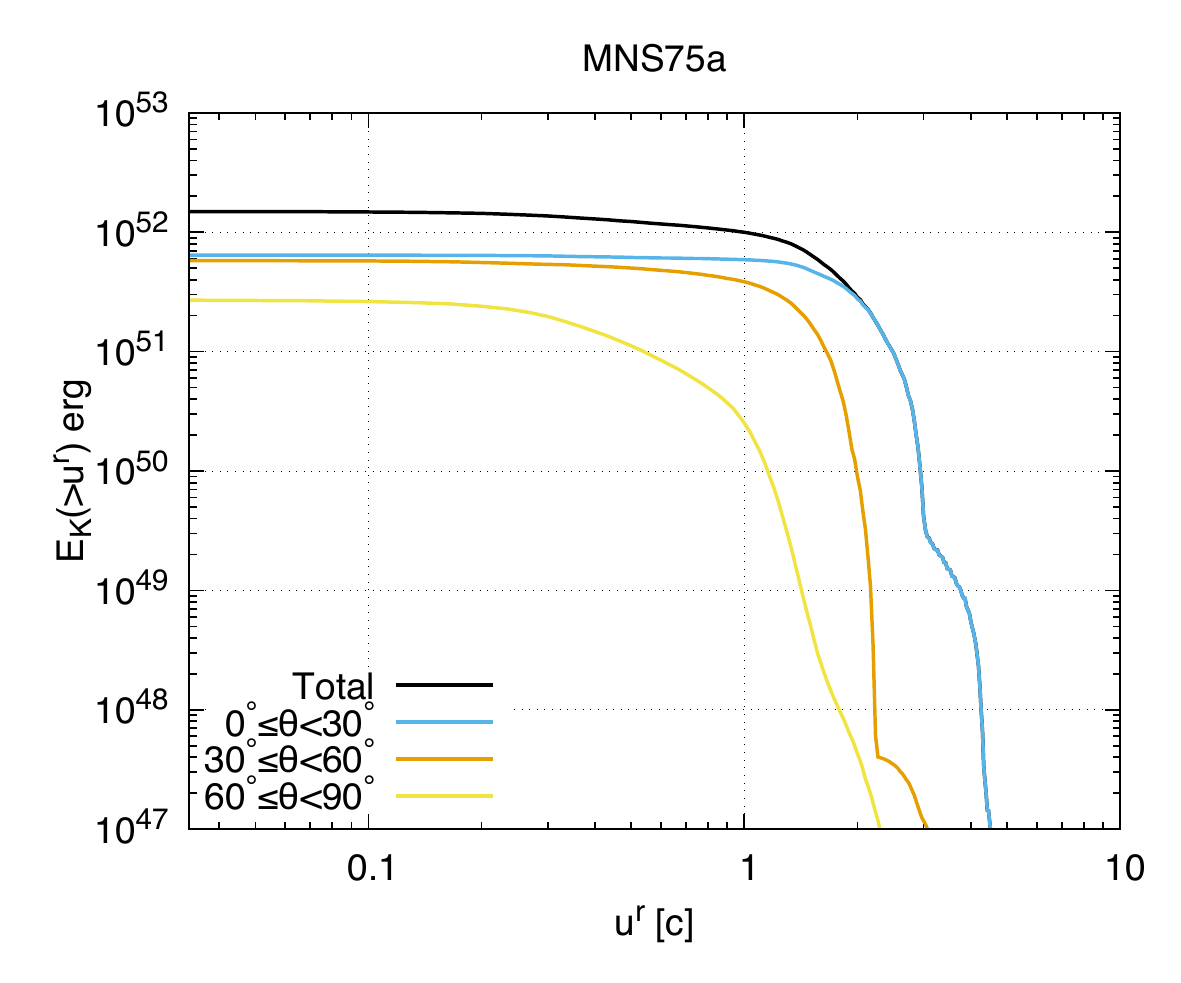}
 	 \caption{Upper panel: The velocity distribution of the ejecta at $t=0.1\,{\rm d}$ for all the models employed in this paper. Lower panel: The same as the upper panel but for three different polar directions for model MNS75a. }
	 \label{fig:ked}
\end{figure}

Figure~\ref{fig:ked} shows the kinetic energy distributions of ejecta 
at $t=0.1\,{\rm d}$ as a function of the radial four-velocity defined by 
\begin{align}
	E_{\rm K}(>u^r)=\int_{>u^r} \rho _* c^2 (w-1)dV.
\end{align}
 As we already found in Table~\ref{tb:model}, the ejecta become more energetic as the values of $\alpha_{\rm d}$ and/or $\sigma_{\rm c}$ increase. The kinetic energy distributions for the MHD models are composed broadly of two components. One is a trans-relativistic component with the Lorentz factor of $ \alt2$, and the other is a mildly relativistic component with the Lorentz factor of $\approx2$--$5$. The former component dominates the total kinetic energy of the ejecta with $\sim10^{51}\,{\rm erg}$ for $\sigma_{\rm c}=1\times 10^7\,{\rm s}^{-1}$ (MNS70a and MNS70b) and $\agt 10^{52}\,{\rm erg}$ for $\sigma_{\rm c} \geq 3 \times 10^7\,{\rm s}^{-1}$ (MNS75a, MNS75b, and MNS80), respectively. The latter component has $\sim10^{48}$--$10^{49}\,{\rm erg}$, which is by more than $\sim2$ orders of magnitudes smaller than the former components.

The bottom panel of Figure~\ref{fig:ked} presents the angular dependence of the kinetic energy. This shows that the kinetic energy of the ejecta is dominated by the matter in a polar region ($\theta\le60^\circ$). In particular, the mildly relativistic component with $u^r/c\agt3~(v^r/c\agt 0.95)$ are only present in the polar region with $\theta\le30^\circ$. This profile is obtained by the following mechanism: The magneto-centrifugal effect initially induces the acceleration of the post-merger ejecta primarily toward the equatorial direction. However, the straightforward acceleration is blocked by the torus surrounding the MNS. By the interaction with the torus, shocks are formed around the torus surface, and thus, the motion of the post-merger ejecta is changed to the polar direction.

In the viscous model, the kinetic energy distribution of ejecta for $u^r/c > 0.05$ is similar to that only with the dynamical ejecta~\citep{Hotokezaka:2018gmo}. This is due to the fact that the post-merger ejecta have only a minor contribution to the fast tail of the ejecta for the viscous model. This is consistent with the fact that the average velocity of the post-merger ejecta is lower than that for the MHD models.

\subsection{Synchrotron afterglow}\label{sec4.2}
The enhancement of the rest mass and total kinetic energy of the fast velocity component in the MHD models has a significant impact on the flux of the ejecta afterglow, because the light curve depends strongly on the Lorentz factor and kinetic energy of the ejecta \citep[see, e.g.,][]{Nakar2018}. It has been suggested that the afterglow of a BNS merger is expected to be extremely bright if a fraction of the remnant NS's rotational energy $\sim 10^{52}\,{\rm erg}$ is converted to the ejecta kinetic energy \citep{Yu:2013kra,Metzger2014MNRAS,Horesh2016ApJ,Yu:2017syg,Li:2018hzy,Beniamini2021ApJ}. Here we present the synchrotron afterglow light curve arising from the merger ejecta using the kinetic energy distribution shown in Figure \ref{fig:ked}.

The light curves for the synchrotron afterglow are calculated using the same analysis as that in~\cite{Hotokezaka:2018gmo} with  the equipartition parameters at shocks, $\epsilon_e$ (the fraction of the energy which accelerated electrons have) and $\epsilon_B$ (the fraction of the energy which the magnetic field has), are set to be 0.1 and $0.01$, respectively. The energy distribution of the accelerated electrons is assumed to be a power-law distribution of the electron Lorentz factor with the power of $p=2.2$. Note that \cite{Margalit2021ApJ} show that thermal electrons are expected to contribute significantly to the radio flux for mildly relativistic outflows.  

\begin{figure}
 	 \includegraphics[width=1\linewidth]{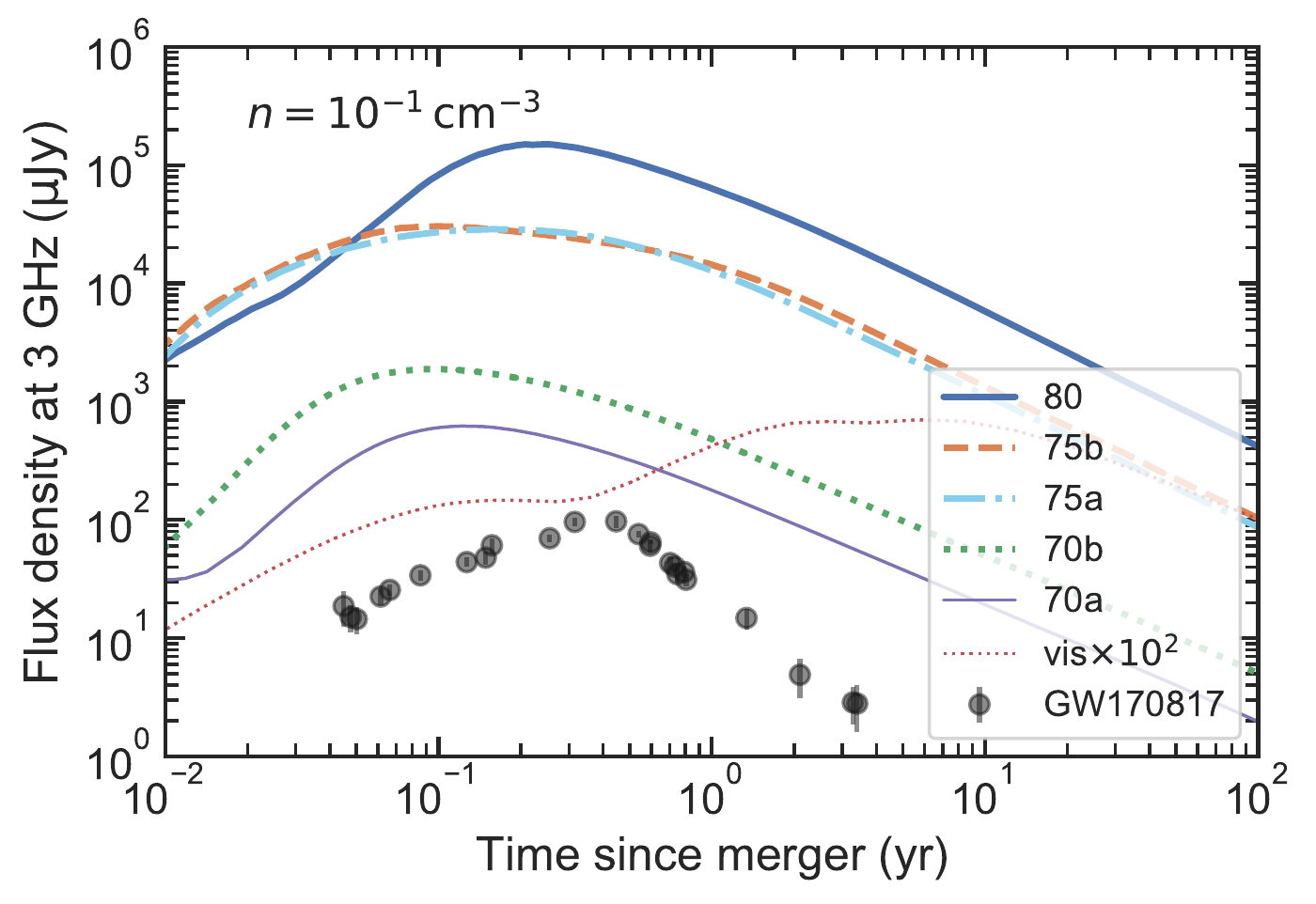}\\
 	 \includegraphics[width=1\linewidth]{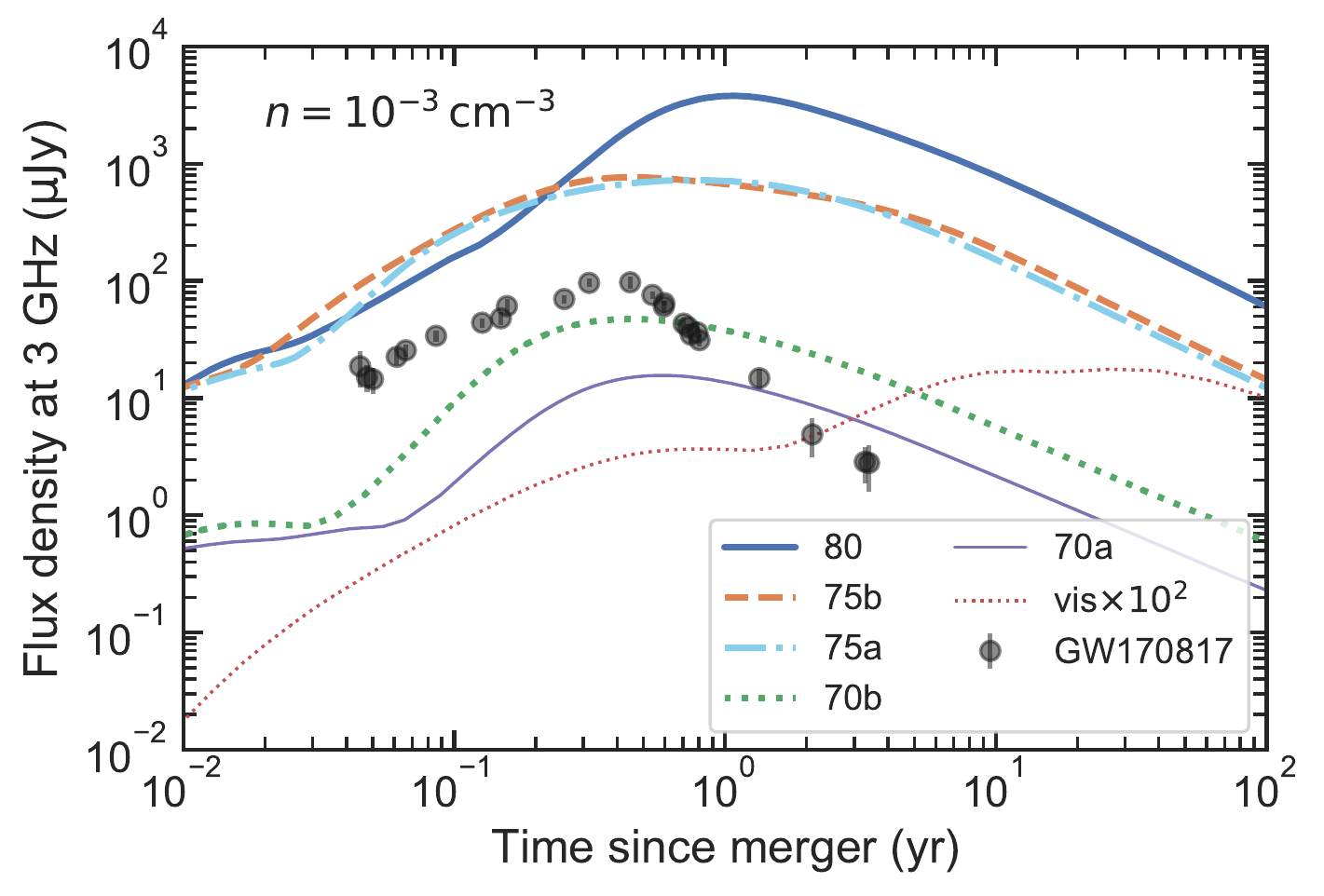}\\
 	 \includegraphics[width=1\linewidth]{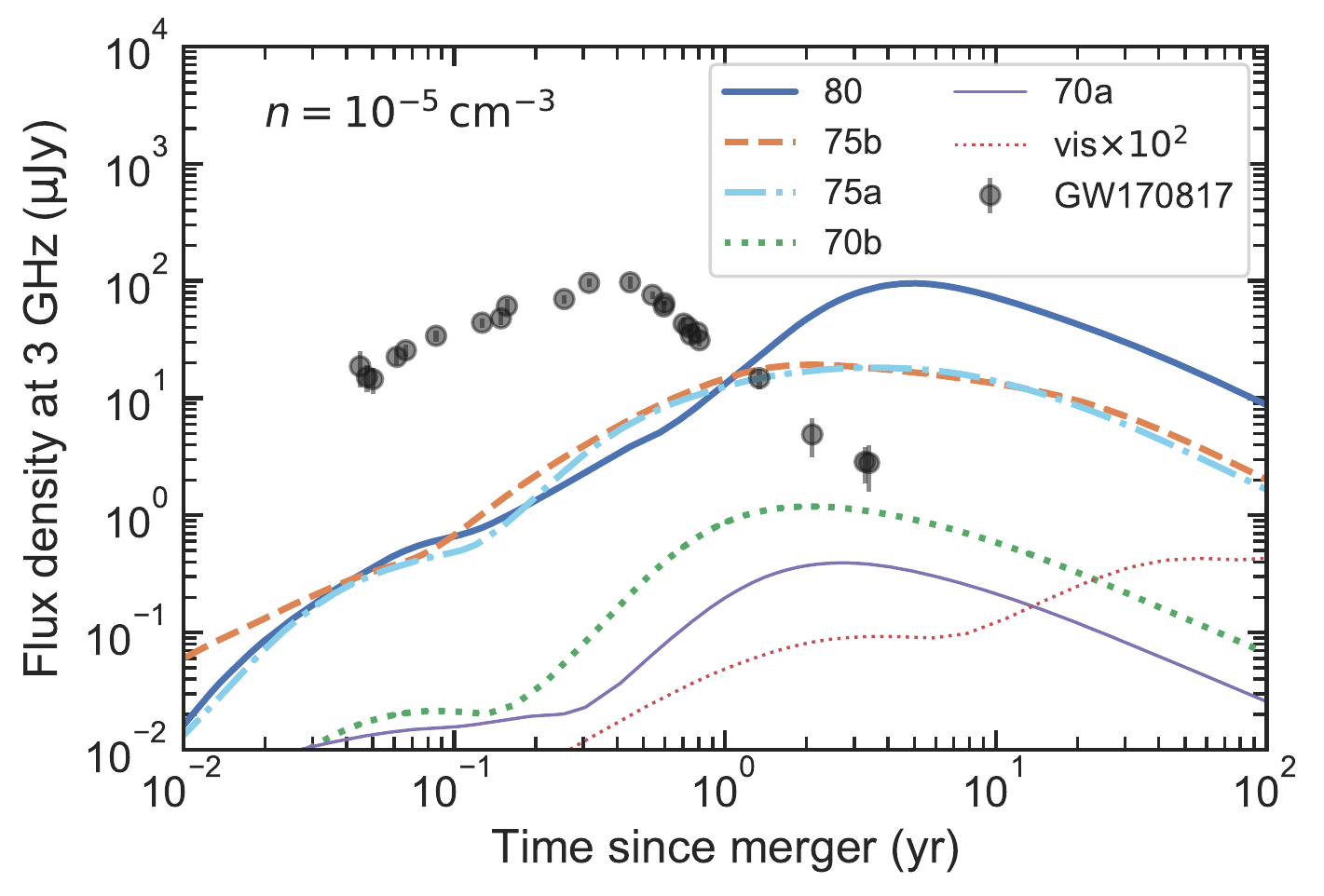}
 	 \caption{The light curves of the kilonova radio afterglow at $3\,{\rm GHz}$ for the MHD models at the distance to the source of $200$\,Mpc.  The top, middle, and bottom panels denote the results for the cases that the ISM density, $n$, is  $10^{-1}$, $10^{-3}$, and $10^{-5}\,{\rm cm^{-3}}$, respectively. Here the parameters of $\epsilon_e=0.1$, $\epsilon_B=0.01$, and $p=2.2$ are used for modeling the shock. We note that the luminosity for the viscous model (``vis$\times10^2$") is scaled-up by a factor of $100$. The data points denote the observation of GW170817 taken from~\cite{Makhathini:2020ece}.}
	 \label{fig:radio}
\end{figure}

Figure~\ref{fig:radio} shows the light curves of the kilonova radio afterglow at $3\,{\rm GHz}$. The top, middle, and bottom panels denote the results for the cases of the different interstellar medium (ISM) densities, $n=10^{-1}$, $10^{-3}$, and $10^{-5}\,{\rm cm^{-3}}$, respectively. Broadly speaking, there are two components in the radio light curves: one is the main component which peaks at $t\geq0.1\,{\rm yr}$ and the other is the faint early-rising component present for $t\leq0.1\,{\rm yr}$. The emission around the peak luminosity arises from the trans-relativistic ejecta component with the Lorentz factor of $ \alt2$. The mildly relativistic component with the Lorentz factor of $\approx3$--$5$ contributes to the early radio light curves for $t\leq0.1\,{\rm yr}$. The contribution of the latter component to the total radio luminosity is minor compared to the former component, and hence, we focus mainly on the former one in the following, although the latter could be important for the follow-up observation in the early phase. 

For either of the cases, the radio emission around the peak time in the MHD models is much brighter than that predicted by the model only with the dynamical ejecta~\citep{Hotokezaka:2018gmo} and by the viscous model. The peak luminosity   varies by more than two orders of magnitude for a fixed ISM density, and the larger luminosities are achieved for the larger values of $\sigma_{\rm c}$ and/or $\alpha_{\rm d}$. Notably, the peak flux of the models with $\sigma_{\rm c}=1\times10^{8}\,{\rm s^{-1}}$ exceeds $\sim 0.1\,{\rm mJy}$ even for the distance to the source of 200\,Mpc and a very low density $n\alt 10^{-5}\,{\rm cm^{-3}}$. This peak flux is as bright as the radio afterglow peak observed in GW170817. On the other hand, $n\ge10^{-3}\,{\rm cm^{-3}}$ is required for the radio emission to be brighter than $\sim 0.1\,{\rm mJy}$ for the models with $\sigma_{\rm c}=1\times10^{7}\,{\rm s^{-1}}$. This suggests that the radio afterglow will carry important information for the magnetic-field enhancement in the merger remnants with a long-lived MNS.

The time of the peak luminosity also depends on the ISM density and the dynamo parameters. In particular, the peak time is delayed for the models with large values of $\sigma_{\rm c}$. This reflects the fact that larger total kinetic energy of the ejecta is achieved for larger values of $\sigma_{\rm c}$ because of the longer dissipation time scale of the magnetic fields (see Table~\ref{tb:model}). Interestingly, the rising part of the radio light curves for the models with $\sigma_{\rm c}\ge3\times10^{7}\,{\rm s^{-1}}$ has a shape similar to each other because of the similarity in the kinetic energy distribution for $u^r/c \gtrsim 2$. Measuring the slope of the early radio light curves may have an important implication to the magnetic field amplification. 

\begin{figure}
 	 \includegraphics[width=1\linewidth]{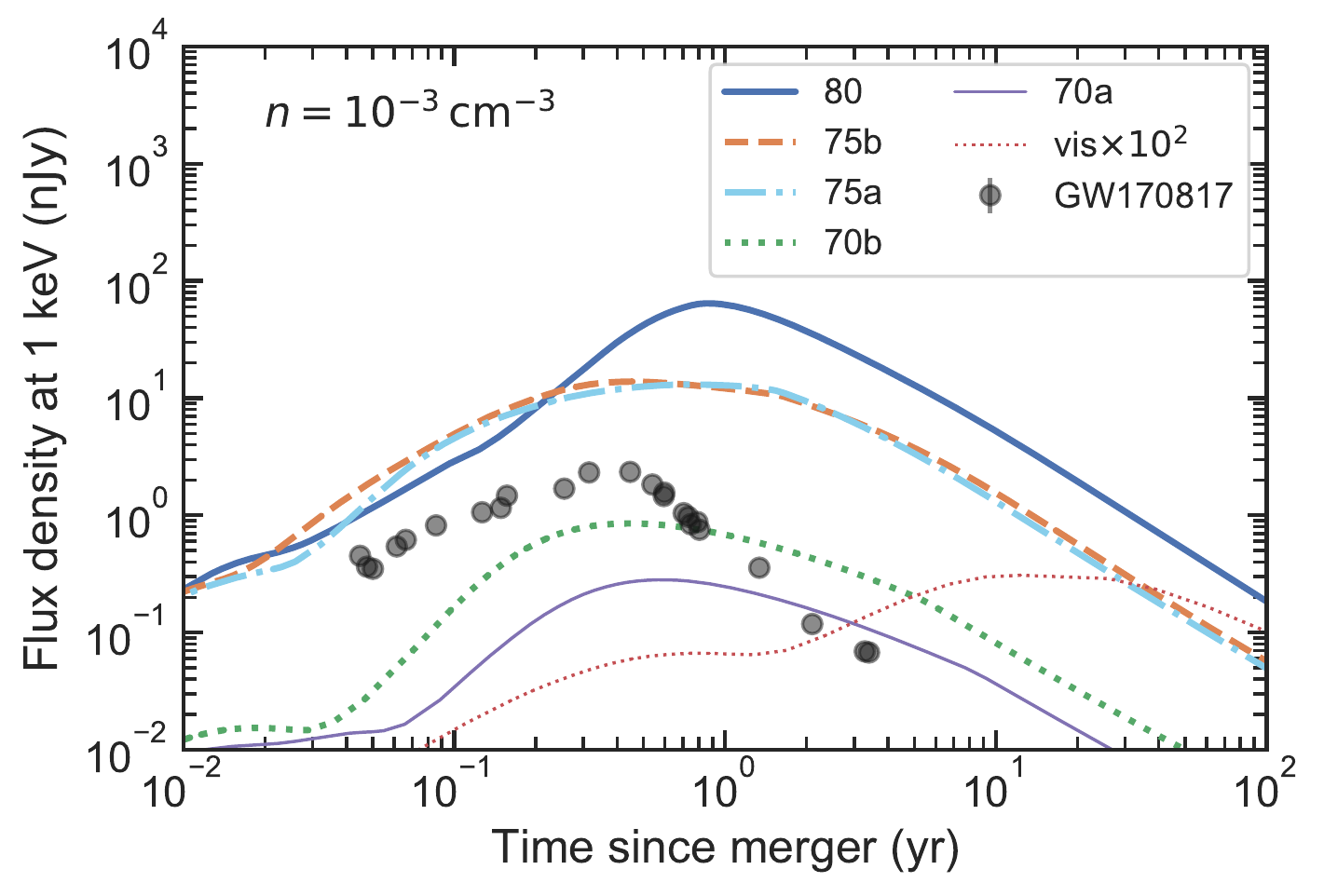}
 	 \caption{The same as Figure~\ref{fig:radio} but for that in an X-ray band (1~keV). The results for the ISM density, $n=10^{-3}\,{\rm cm^{-3}}$  is shown. The data points denote the observation for GW170817 obtained by extrapolation from the radio-band observation in~\cite{Makhathini:2020ece} assuming a power-law frequency dependence with the power of $-0.584$ for the emission.}
	 \label{fig:xray}
\end{figure}

Figure~\ref{fig:xray} shows the same as Figure~\ref{fig:radio} but for that in an X-ray band (1~keV). Here, the case of $n=10^{-3}\,{\rm cm^{-3}}$ is shown. In Figure~\ref{fig:xray}, we also plot the data points for GW170817 obtained by extrapolation from the data at $3$ GHz assuming a single power law spectrum with the best fit index of $-0.584$~\citep{Makhathini:2020ece}.  As is the same in the radio band, the X-ray emission for the models with $\sigma_{\rm c}\ge3\times10^{7}\,{\rm s^{-1}}$ is significantly brighter than that observed in GW170817 for $n=10^{-3}\,{\rm cm^{-3}}$. The X-ray light curves become fainter than the flux extrapolated from the radio light curve with $\nu^{-(p-1)/2}$ once the cooling frequency crosses 1 keV. This occurs around the peak time for $\sigma_{\rm c}\ge3\times10^{7}\,{\rm s^{-1}}$ and around $t=10$\,yrs for $\sigma_{\rm c}=1\times 10^{7}\,{\rm s^{-1}}$. This time scale is shorter for higher ISM densities, e.g., a week for $\sigma_{\rm c}\ge3\times10^{7}\,{\rm s^{-1}}$ at $n=10^{-1}\,{\rm cm^{-3}}$. By contrast, the X-ray luminosity from the ejecta is not as high as that of GW170817 for $\sigma_{\rm c}=1\times 10^7\,{\rm s}^{-1}$ (i.e., for the short dissipation time scale) and the viscous model.

\subsection{Kilonovae}\label{sec4.3}

\begin{figure*}
 	 \includegraphics[width=0.45\linewidth]{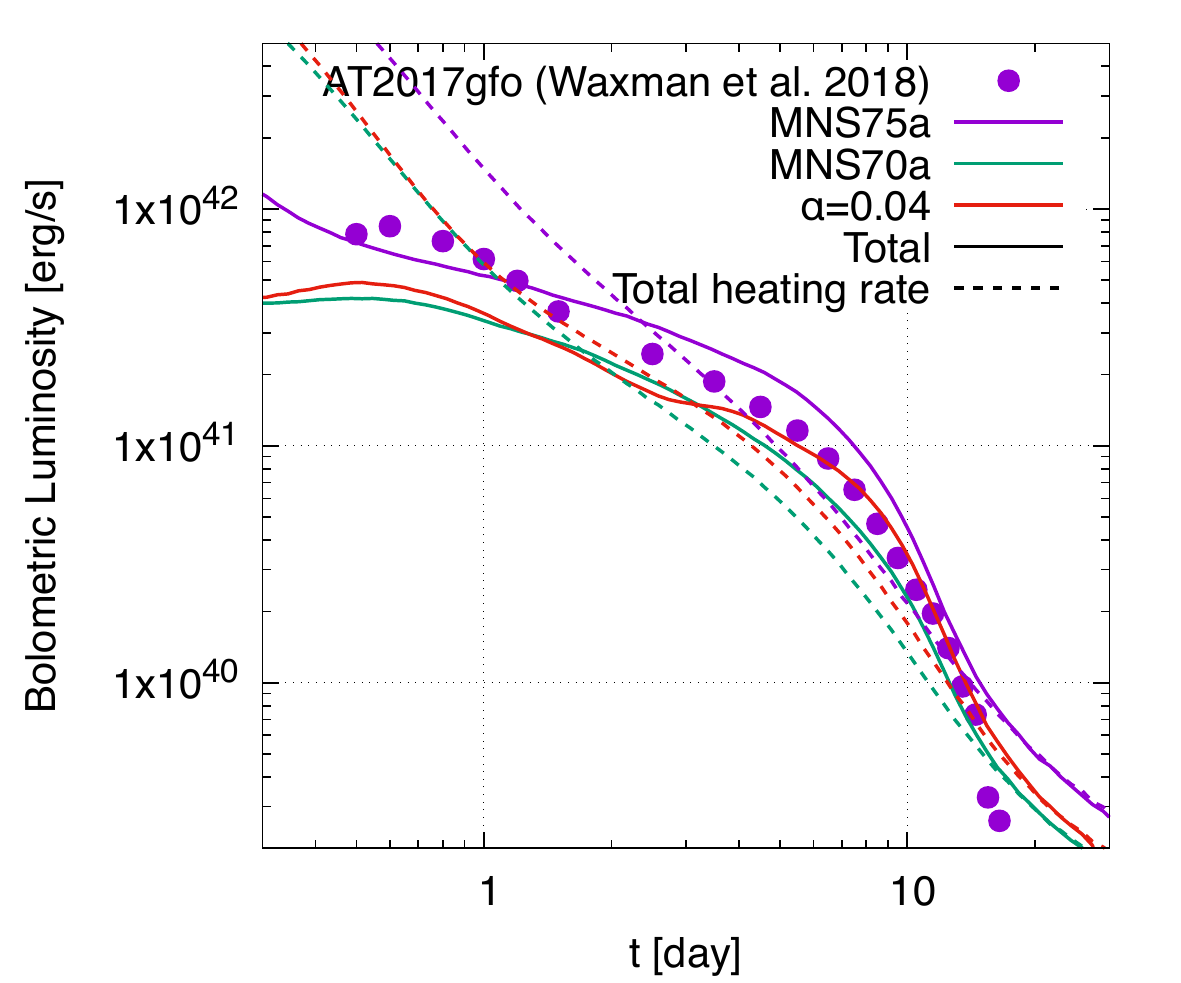}
 	 \includegraphics[width=0.45\linewidth]{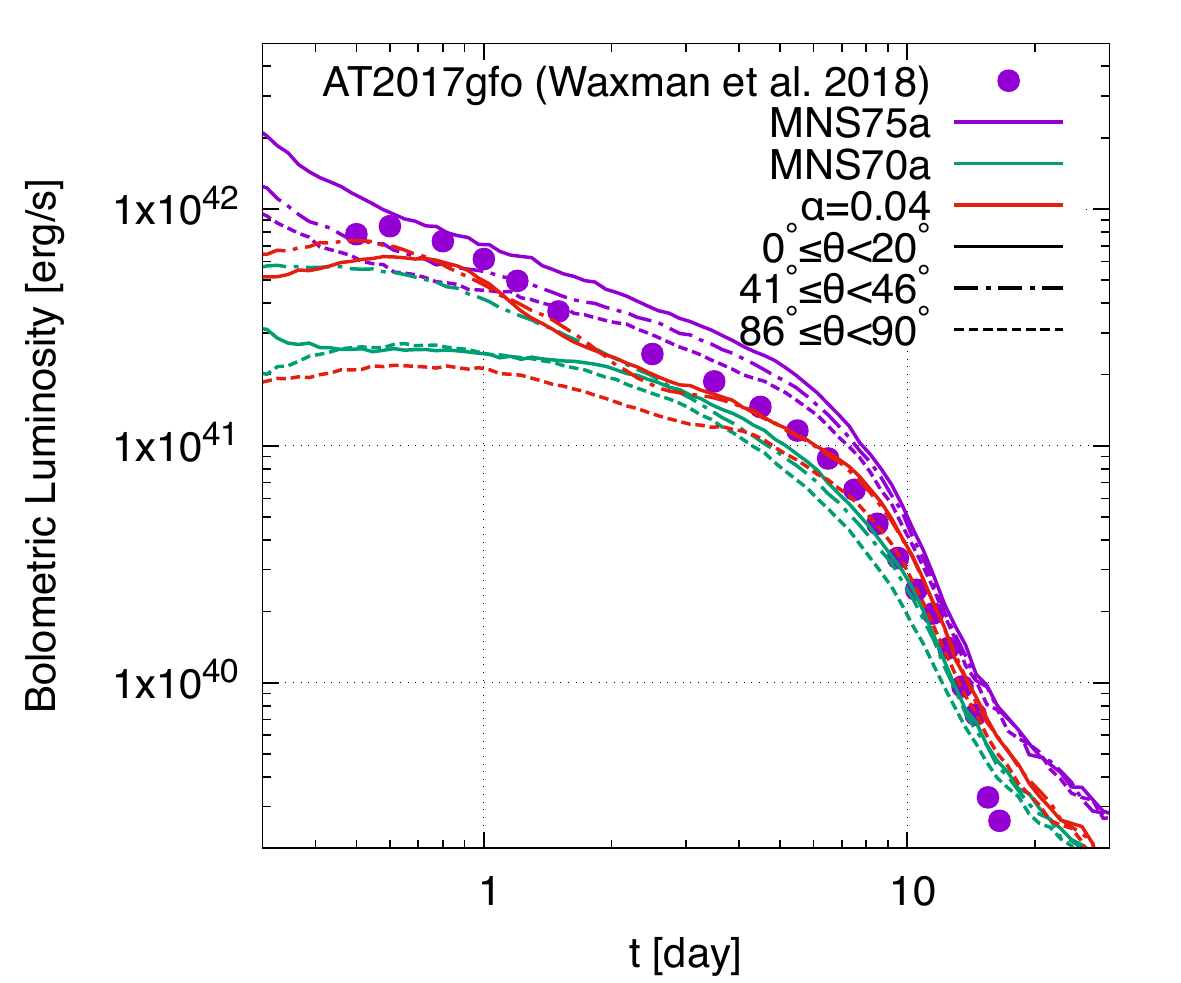}
 	 \caption{The bolometric light curves for models MNS75a (purple curves), MNS70a (green curves), and the viscous model ($\alpha=0.04$; red curves). The left panel shows the total bolometric luminosity (solid curves) and total heating rate (dashed curves). The solid, dotted, and dashed curves in the right panel show the isotropic-equivalent bolometric luminosity observed from $0^\circ\leq\theta\leq20^\circ$, $41^\circ\leq\theta\leq46^\circ$, and $86^\circ\leq\theta\leq90^\circ$, respectively. The bolometric luminosity observed in GW170817 is shown by the filled circles using the data in~\cite{Waxman:2017sqv}}
	 \label{fig:bol}
\end{figure*}

The light curves of kilonovae are calculated for models MNS80, MNS75a, MNS70a, and viscous model using the same method as in our previous paper~\citep{Kawaguchi:2020vbf}, i.e.,  using a wavelength-dependent radiative transfer simulation code~\citep{Tanaka:2013ana,Tanaka:2017qxj,Tanaka:2017lxb,Kawaguchi:2019nju,Kawaguchi:2020vbf}. In this code, the photon transfer is simulated by a Monte Carlo method for given ejecta profiles composed of the density, velocity, and element abundance. The time-dependent thermalization efficiency is taken into account following an analytic formula derived by~\cite{Barnes:2016umi}. The ionization and excitation states are determined under the assumption of the local thermodynamic equilibrium (LTE) by using the Saha ionization and Boltzmann excitation equations. We note, however, that the assumption of the LTE is not always valid in the low density region of the ejecta for the later epoch. We discuss the possible non-LTE effects to the kilonova light curves in Section~\ref{sec:nonLTE}.

For the photon-matter interaction, bound-bound, bound-free, and free-free transitions and electron scattering are taken into account for the transfer of optical and infrared photons~\citep{Tanaka:2013ana,Tanaka:2017qxj,Tanaka:2017lxb}. The formalism of the expansion opacity~\citep{1983ApJ...272..259F,1993ApJ...412..731E,Kasen:2006ce} and the updated line list derived in~\cite{Tanaka:2019iqp} are employed for the bound-bound transitions. The new line list is constructed by an atomic structure calculation for the elements from $Z=26$ to $Z=92$~(see \cite{Tanaka:2019iqp} for details), and supplemented by the Kurucz's line list for $Z < 26$~\citep{1995all..book.....K}.

The radiative transfer simulations are performed from $t=0.1\,{\rm d}$ to $30\,{\rm d}$. As the background hydrodynamical state, the density profiles of the HD simulations at $t=0.1\,{\rm d}$ are used with the assumption of the homologous expansion. The initial internal energy and temperature for the radiative transfer simulations are also determined from those obtained by the HD simulations. The spatial distributions of the heating rate and element abundances are determined by the table obtained by the nucleosynthesis calculations referring to the injected time and angle of the fluid elements. Note that the element abundances at $t=1\,{\rm d}$ (the right panel in Fig.~\ref{fig:nucleosynthesis}) are used during the entire time evolution in the radiative transfer simulations to reduce the computational cost, because this simplified prescription (i.e., neglecting the effects that come from the late-time nucleosynthesis) gives an only minor systematic error on the numerical results.

We first focus on models MNS75a, MNS70a, and viscous model. Figure~\ref{fig:bol} shows the bolometric luminosity as a function of time for models MNS75a, MNS70a, and the viscous model ($\alpha=0.04$). The bolometric luminosity for MNS75a is always larger than those for MNS70a and the viscous model. This is primarily due to the larger ejecta mass of MNS75a than those of MNS70a and the viscous model, but the higher total specific heating rate and thermalization efficiency are also relevant for the difference. The release of the internal energy stored in the matter until the initial period also contributes to the total bolometric luminosity for $t\le0.5\,{\rm d}$. 

For a given model, the bolometric luminosity becomes larger for smaller viewing angles (except for that of model MNS70a observed from $0^\circ\leq\theta\leq20^\circ$ as we discuss later). This is primarily due to the presence of the lanthanide-rich dynamical ejecta in the equatorial region, of which a high opacity plays an important role to block the optical photons coming from the inner dense region (known as the lanthanide curtain effect;~\citealt{Kasen:2014toa,Kawaguchi:2019nju,Bulla:2019muo,Zhu:2020inc,Darbha:2020lhz,Korobkin:2020spe}). The viewing angle dependence is weaker for model MNS75a than those for model MNS70a and the viscous model, because the lanthanide-rich dynamical ejecta are confined more prominently in the equatorial region for model MNS75a than for the other models (cf. Figure~\ref{fig:prof}). 

Model MNS70a and the viscous model show similar total bolometric light curves. This is also the case for the bolometric light curves observed from several viewing angles (see the right panel of Figure~\ref{fig:bol}), except for the face-on observation: The bolometric luminosity observed from $0^\circ\leq\theta\leq20^\circ$ is as small as that observed from $86^\circ\leq\theta\leq90^\circ$ for model MNS70a for $t\lesssim1\,{\rm d}$. This is due to the dominance of the 1st $r$-process peak elements (with a small number of relevant radioactive isotopes) in the polar region. As found in Figure~\ref{fig:prof}, the matter is blown up with high velocity in the MHD models. In addition, appreciable amounts of Y and Zr are present in the matter blown up with high velocity (see Table~\ref{tb:model} for the Y+Zr mass fraction in the ejecta). Note that Y and Zr (particularly, those neutral and the first ionized atoms) have significant contribution to the opacity in the optical wavelength~\citep{Tanaka:2019iqp,Kawaguchi:2020vbf,Ristic:2021ksz}. Photons that diffuse out toward the polar direction have to pass through such a region because Y and Zr are dominantly produced in the high-density part of the ejecta ($v/c\lesssim 0.3$). As a consequence, the emission toward the polar direction is suppressed for model MNS70a in comparison with the viscous model. We note that a small fraction of Sr and lanthanide+actinide  elements that are synthesized and blown up in the polar region also contributes to suppressing the optical emission. By contrast, such suppression of the bolometric luminosity is not very significant for model MNS75a. Our interpretation for this is that the release of the internal energy from the ejecta in the high velocity region, for which the opacity is relatively low, dominates the emission in the early epoch ($t\le0.5\,{\rm d}$).

Note that model MNS70a and the viscous model show broadly similar light curves regardless of the difference of the total ejecta mass. This is because the difference of the total ejecta mass between the two models is due to the difference in the ejecta density in the region where the velocity is less than $0.05\,c$, and such a difference has only a minor effect on the light curves up to two weeks, particularly for the peak flux in the optical wavelengths. 

\begin{figure*}
 	 \includegraphics[width=.5\linewidth]{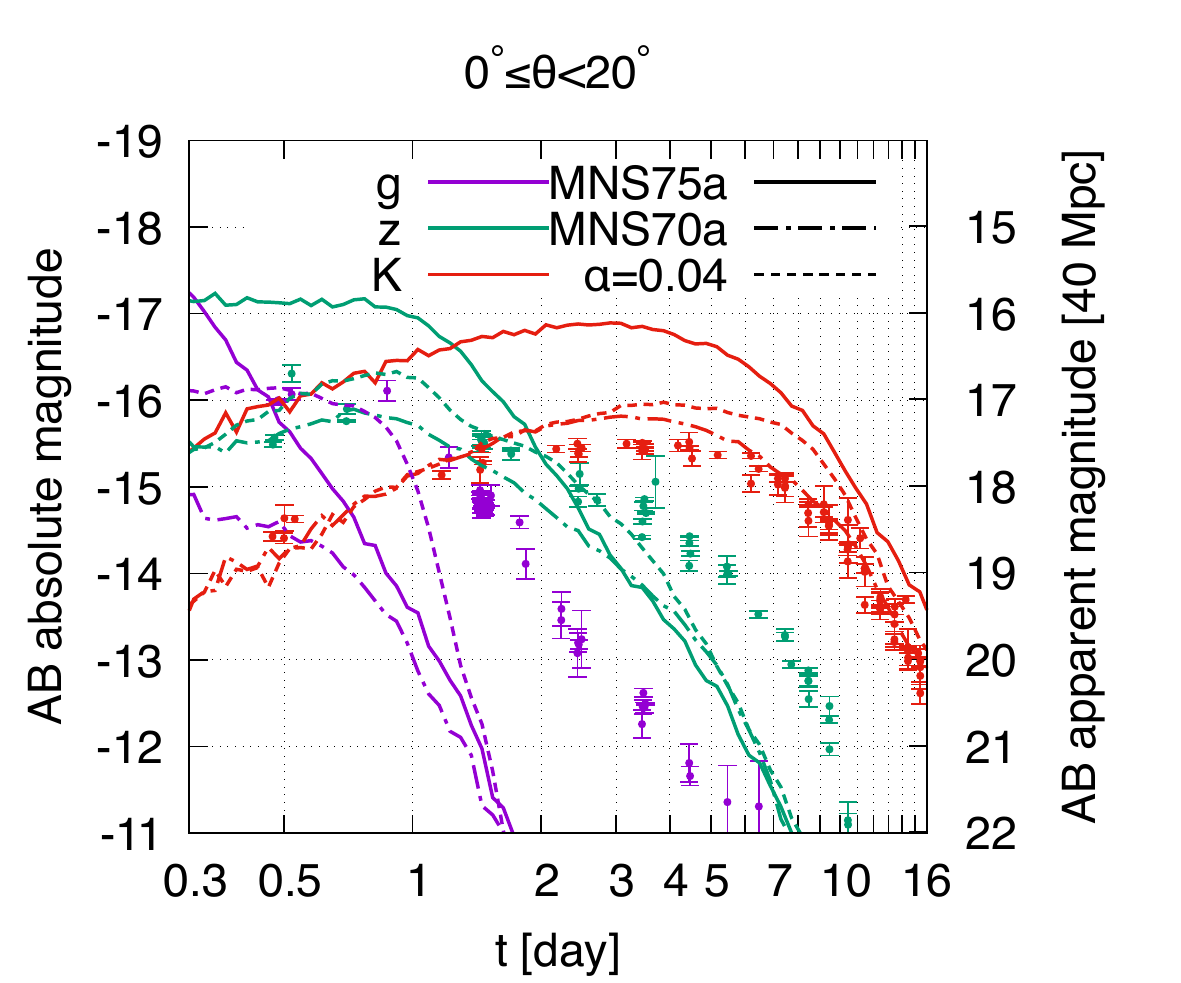}
 	 \includegraphics[width=.5\linewidth]{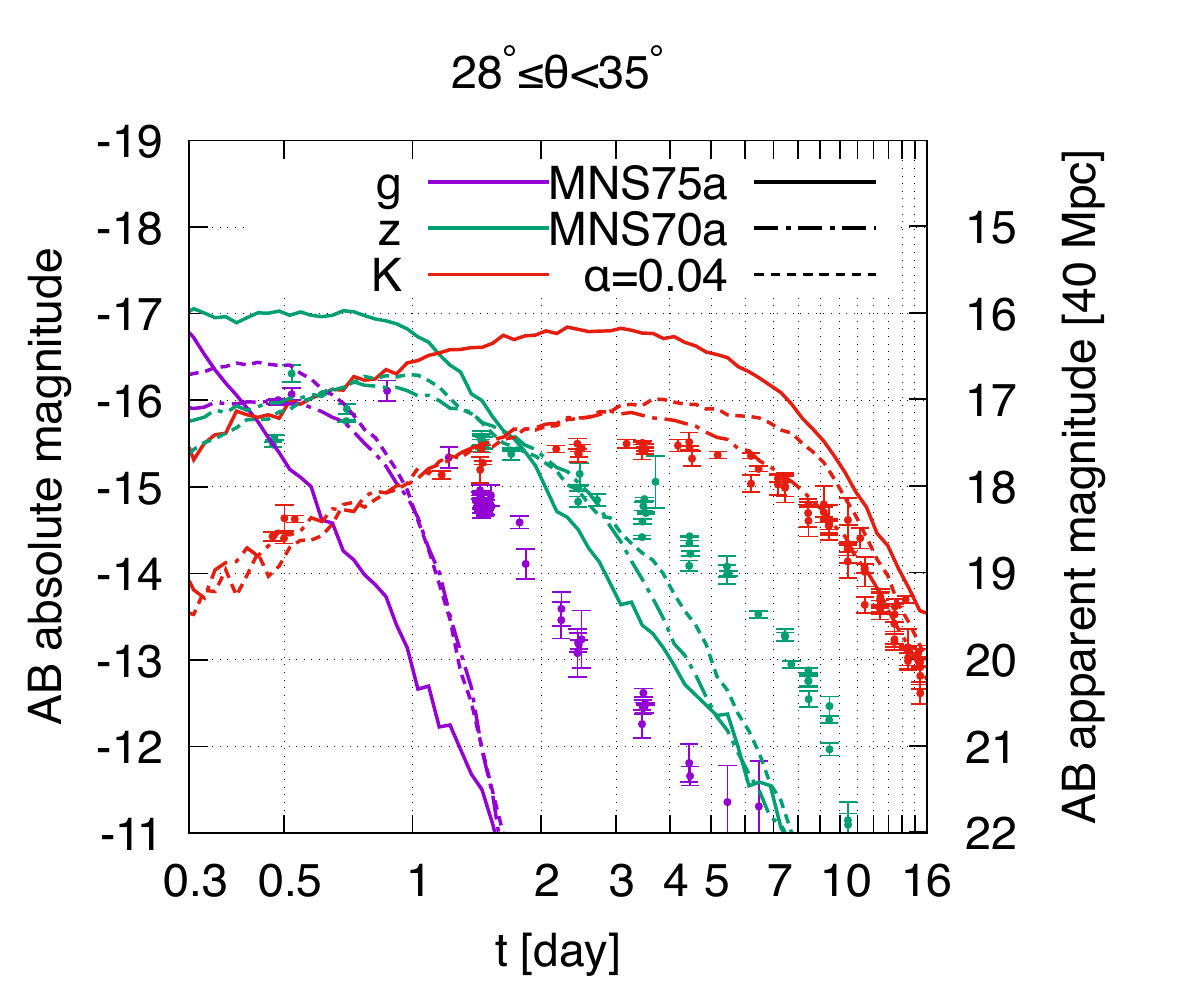}\\
 	 \includegraphics[width=.5\linewidth]{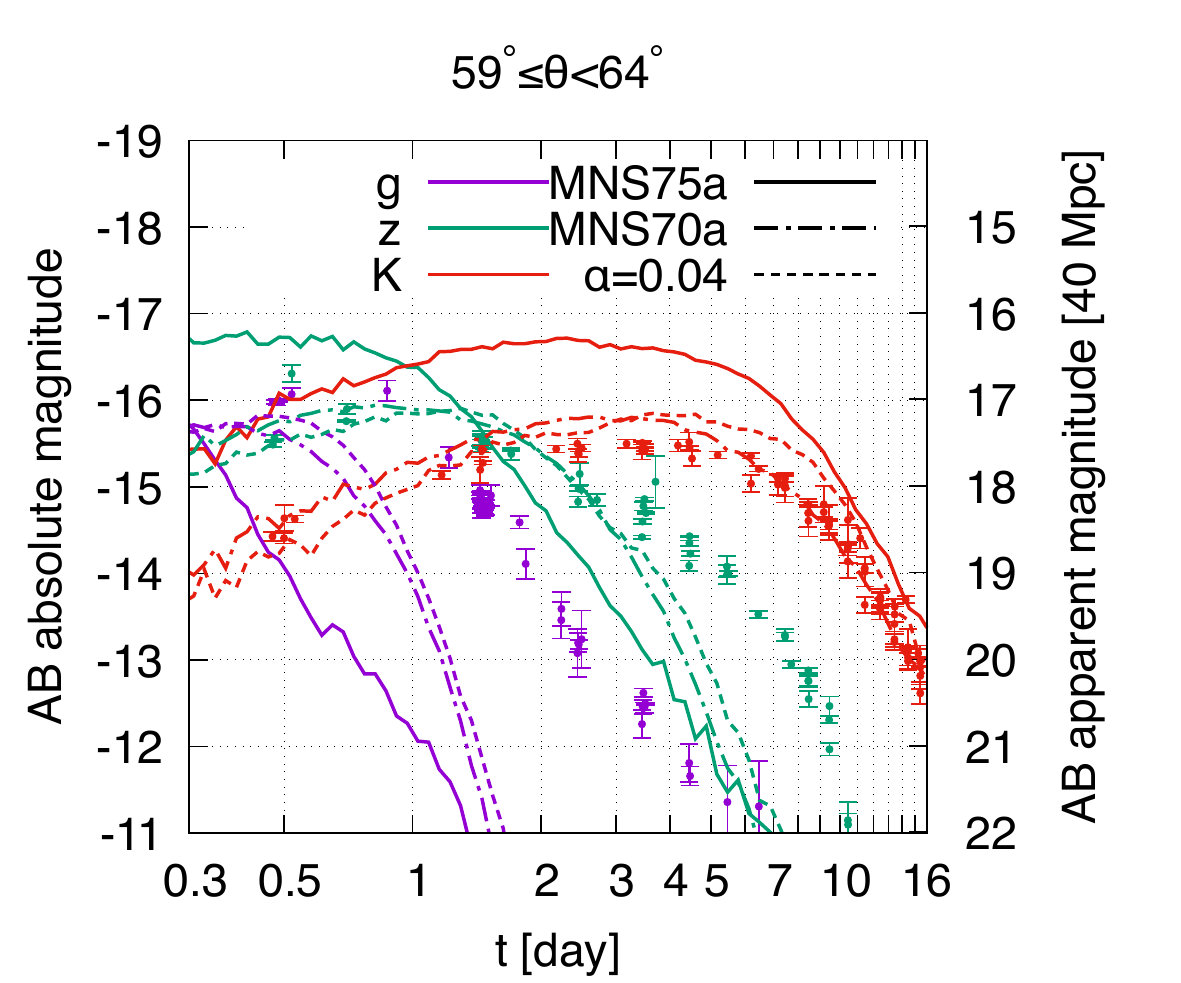}
 	 \includegraphics[width=.5\linewidth]{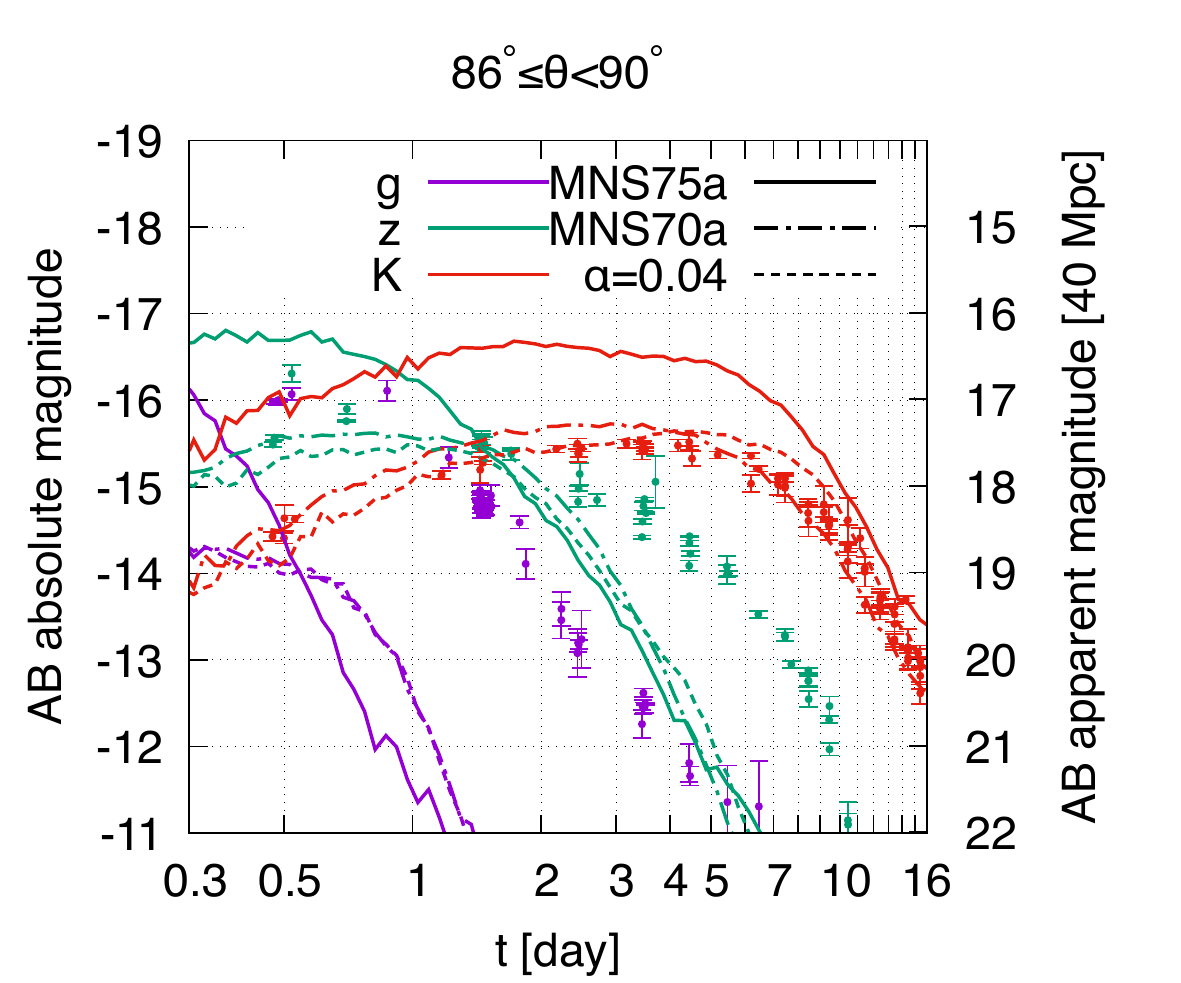}
 	 \caption{{\it gzK}-band light curves for models MNS75a (solid curves), MNS70a (dash-dot curves), and the viscous model ($\alpha=0.04$; short-dash curves). The top left, top right, bottom left, and bottom right panels denote the light curves observed from $0^\circ\leq\theta\leq20^\circ$, $28^\circ\leq\theta\leq35^\circ$, $59^\circ\leq\theta\leq64^\circ$, and $86^\circ\leq\theta\leq90^\circ$, respectively. The data points denote the observation data of GW170817 taken from~\cite{Villar:2017wcc}.} 
	 \label{fig:mag}
\end{figure*}

Figure~\ref{fig:mag} shows the {\it gzK}-band light curves for models MNS75a, MNS70a, and the viscous model at a hypothetical distance to the source of 40\,Mpc. In the early epoch (for $t\le0.5\,{\rm d}$), the kilonova emission for MNS75a is brighter than or as bright as that observed in GW170817. However, the optical ({\it gri}-band) emission fades rapidly, and becomes fainter than that observed in GW170817 for $t\geq 0.5$--$1\,{\rm d}$. On the other hand, the NIR ({\it JHK}-band) emission is always brighter than that observed in GW170817. Similar features have been found in our previous study for a model in which the post-merger ejecta is accelerated to high velocity by a hypothetical remnant activity (see the SMNS case in \citealt{Kawaguchi:2019nju}).

The rapid fading of the optical emission for model MNS75a stems primarily from the rapid exhaustion of the internal energy deposited at the time of ejecta formation. The ejecta optical thickness decreases due to the rapid outgoing expansion of the ejecta for $t\lesssim0.3\,{\rm d}$, and the release of the internal energy stored by the radioactive heating (and also that deposited at the time of ejecta formation) dominates the emission. For $t\lesssim1\,{\rm d}$, such internal energy has already been mostly exhausted by the adiabatic expansion, and as a consequence, the optical emission fades rapidly. The large value of opacity in the optical wavelength due to the presence of the 1st $r$-process peak elements  (including Y and Zr) in the polar region also plays an important role. In particular, the neutral and the first ionized atoms, of which the fraction increases for $t\ge1\,{\rm d}$, have great contributions to the opacity in the optical wavelengths. 

As found in Figure~\ref{fig:bol}, the viewing angle dependence of the {\it gzK}-band emission for model MNS75a is weaker than for the other models, reflecting the approximately spherical ejecta profile and the confinement of the lanthanide-rich dynamical ejecta component in the equatorial region. This feature is advantageous for the observation of the kilonovae from the off-axis directions. 

Model MNS70a and the viscous model show approximately identical light curves, except for those observed from the polar direction ($0^\circ\leq\theta\leq20^\circ$), reflecting the similar ejecta profile. As we described already, the difference in the polar light curves is primarily due to the different density structure and the different fractions of the 1st $r$-process peak elements in the polar region. The similarity in the light curves for the low-$\sigma_{\rm c}$ MHD models and the viscous model implies that the results of viscous hydrodynamics simulations can provide a good phenomenological model for the first-principle MHD model, in the case that the intrinsic MHD effects such as the magneto-centrifugal effect~\citep{BP82} are not very strong.

Kilonova light curves for model MNS70a and viscous model are in a fair agreement with those for GW170817. This suggests that after the merger of the BNS in GW170817, a long-lived MNS with an insufficient magnetic-field amplification or with a short dissipation time scale of the amplified magnetic field (shorter than the mass ejection time scale) might be formed. The weak magnetic-field effect is consistent with the non-detection of bright radio waves associated with the fast ejecta component. However, as we already showed in Figure~\ref{fig:nucleosynthesis}, the abundance patterns of the $r$-process elements derived for model MNS70a and viscous model do not agree with the solar-abundance pattern. This suggests that GW170817 might be the rare event of the BNS mergers, under the hypothesis that the BNS mergers are the major site for the $r$-process nucleosynthesis and the solar abundance pattern is universal in the universe (see also the next subsection for the current uncertainty).

\begin{figure}
 	 \includegraphics[width=1\linewidth]{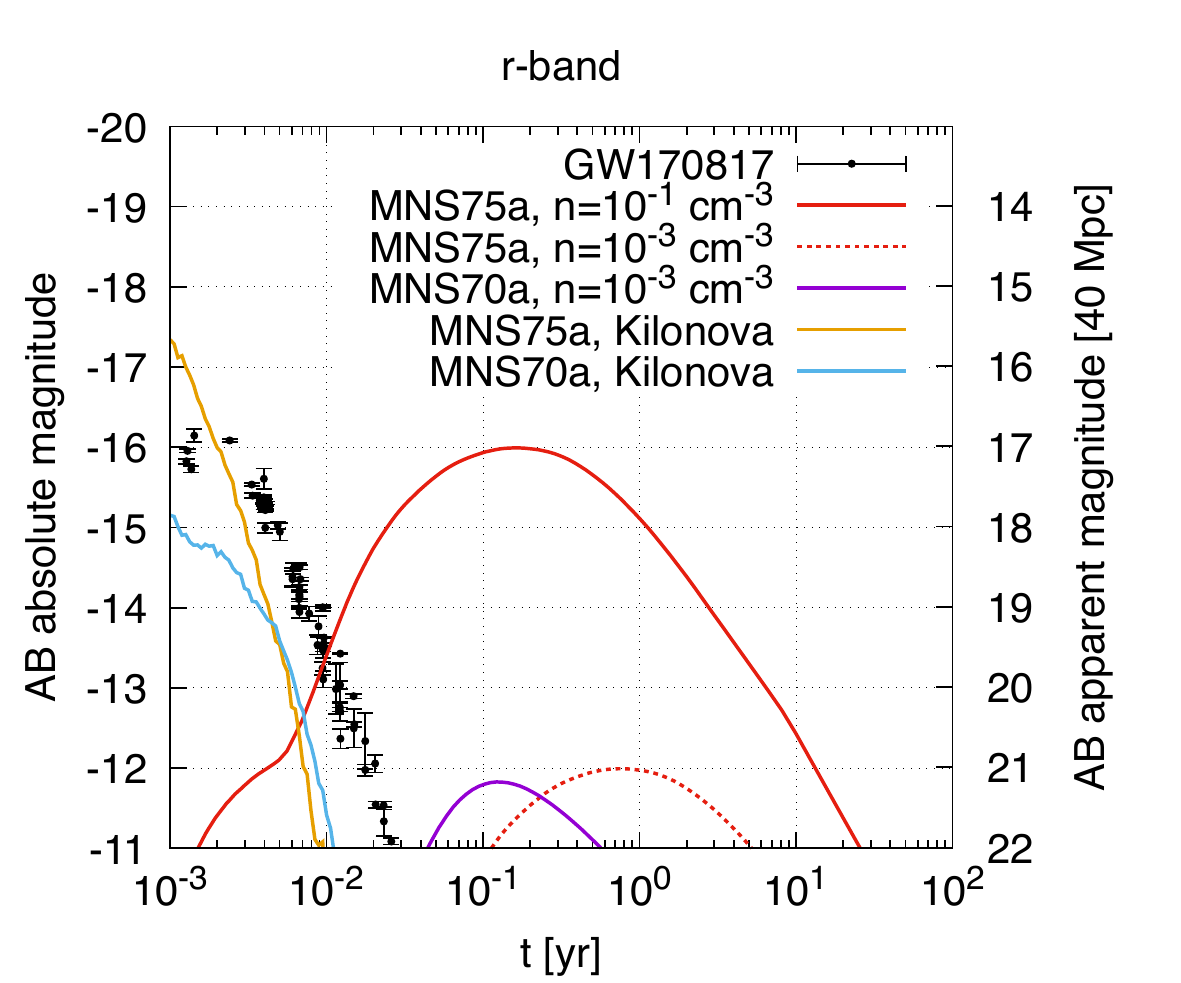}
 	 \caption{Comparison of the kilonova (``MNS75a, kilonova" and ``MNS70a, kilonova") and kilonova afterglow light curves  (``MNS75a, $n=10^{-1}\,{\rm cm}^{-3}$", ``MNS75a, $n=10^{-3}\,{\rm cm}^{-3}$", and ``MNS70a, $n=10^{-3}\,{\rm cm}^{-3}$") in the {\it r}-band. The light curves observed from $0^\circ\leq\theta\leq20^\circ$ are shown for kilonova models. The data points denote the observation of GW170817 in the {\it r}-band taken from~\cite{Villar:2017wcc}.}
	 \label{fig:rband}
\end{figure}

Figure~\ref{fig:rband} compares the kilonova and kilonova afterglow light curves in the {\it r}-band. The {\it r}-band emission from the kilonova afterglow reaches its peak magnitude at $0.1$--$1$ yr. Because the kilonova emission has typically already faded at such an epoch, the afterglow emission can be a characteristic feature for the MHD activity. The afterglow emission could be as bright as the kilonova emission and that of GW170817 intrinsically in the optical wavelength if $\sigma_{\rm c}\ge3\times10^{7}\,{\rm s^{-1}}$ and $n\gtrsim 10^{-1}\,{\rm cm}^{-3}$. For such a case, the emission will be observable even up to the distance of 200 Mpc by 1-m class telescopes~\citep{Nissanke:2012dj}. On the other hand, the afterglow emission will be $\approx21$ and $25$ mag in the {\it r}-band for 40 and 200 Mpc, respectively, if $\sigma_{\rm c}\le3\times10^{7}\,{\rm s^{-1}}$ or $n\lesssim 10^{-3}\,{\rm cm}^{-3}$, and 4/8-m class telescopes are necessary for the follow-up observation.

\begin{figure*}
 	 \includegraphics[width=0.45\linewidth]{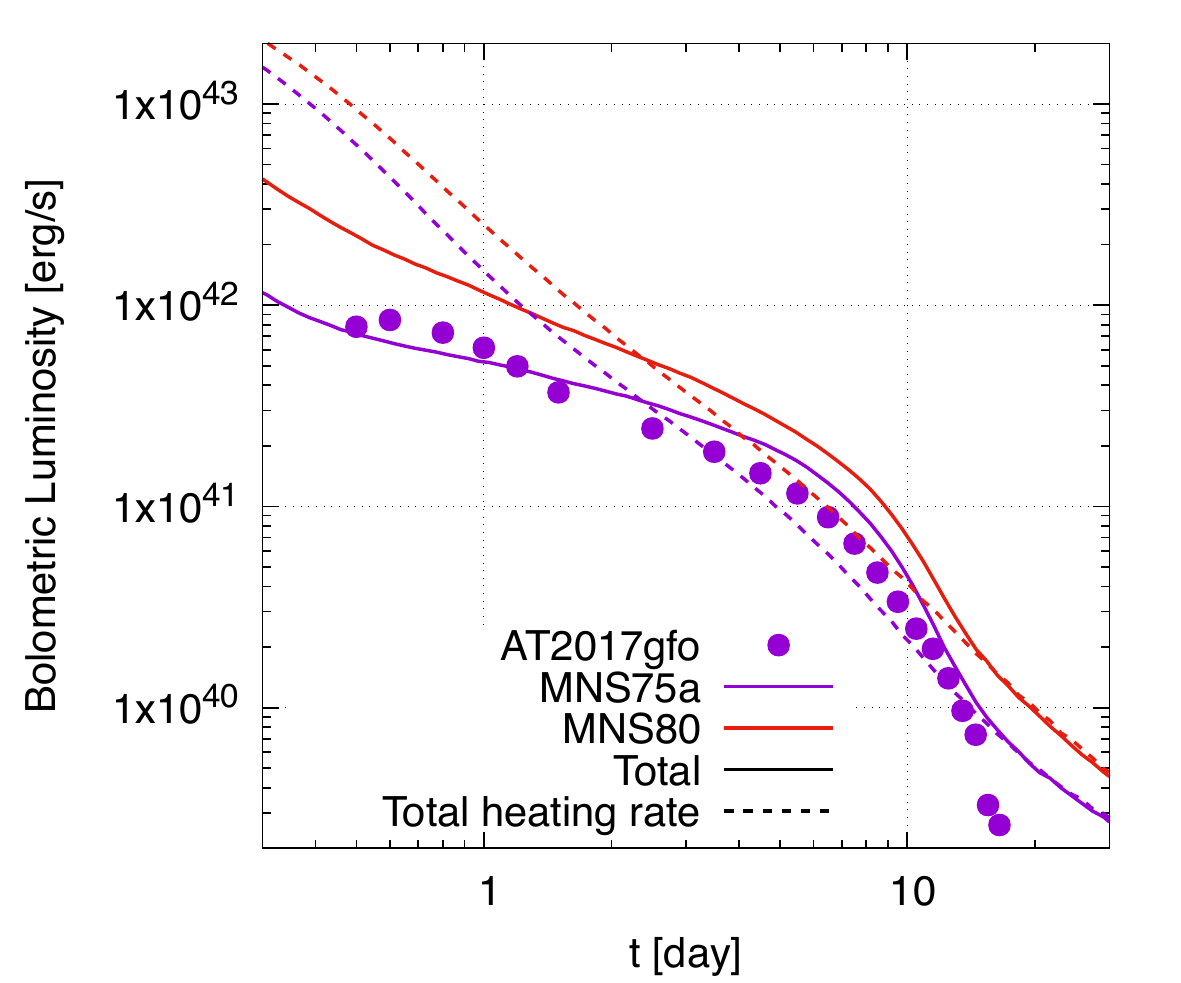}
 	 \includegraphics[width=0.45\linewidth]{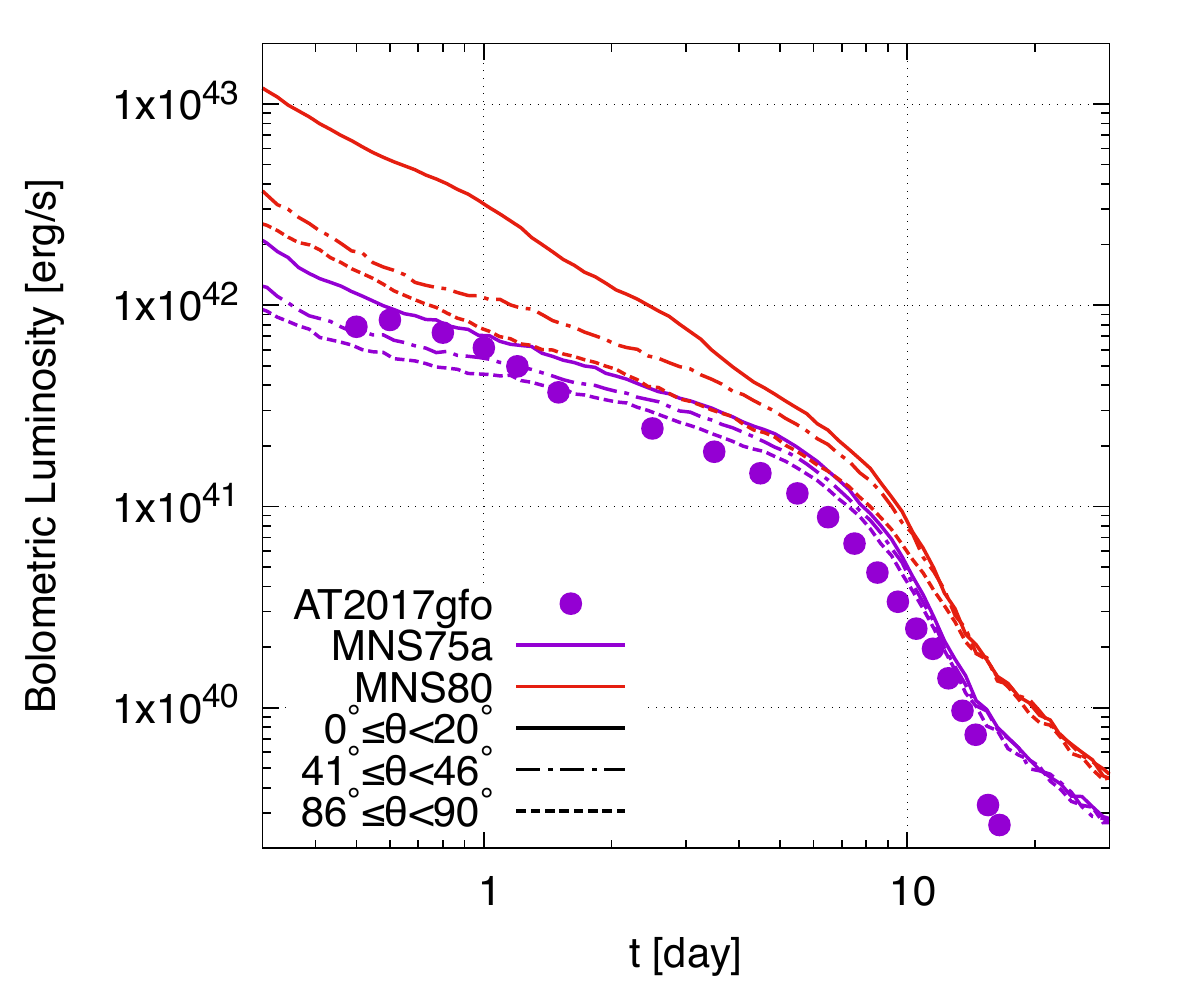}\\
 	 \includegraphics[width=.5\linewidth]{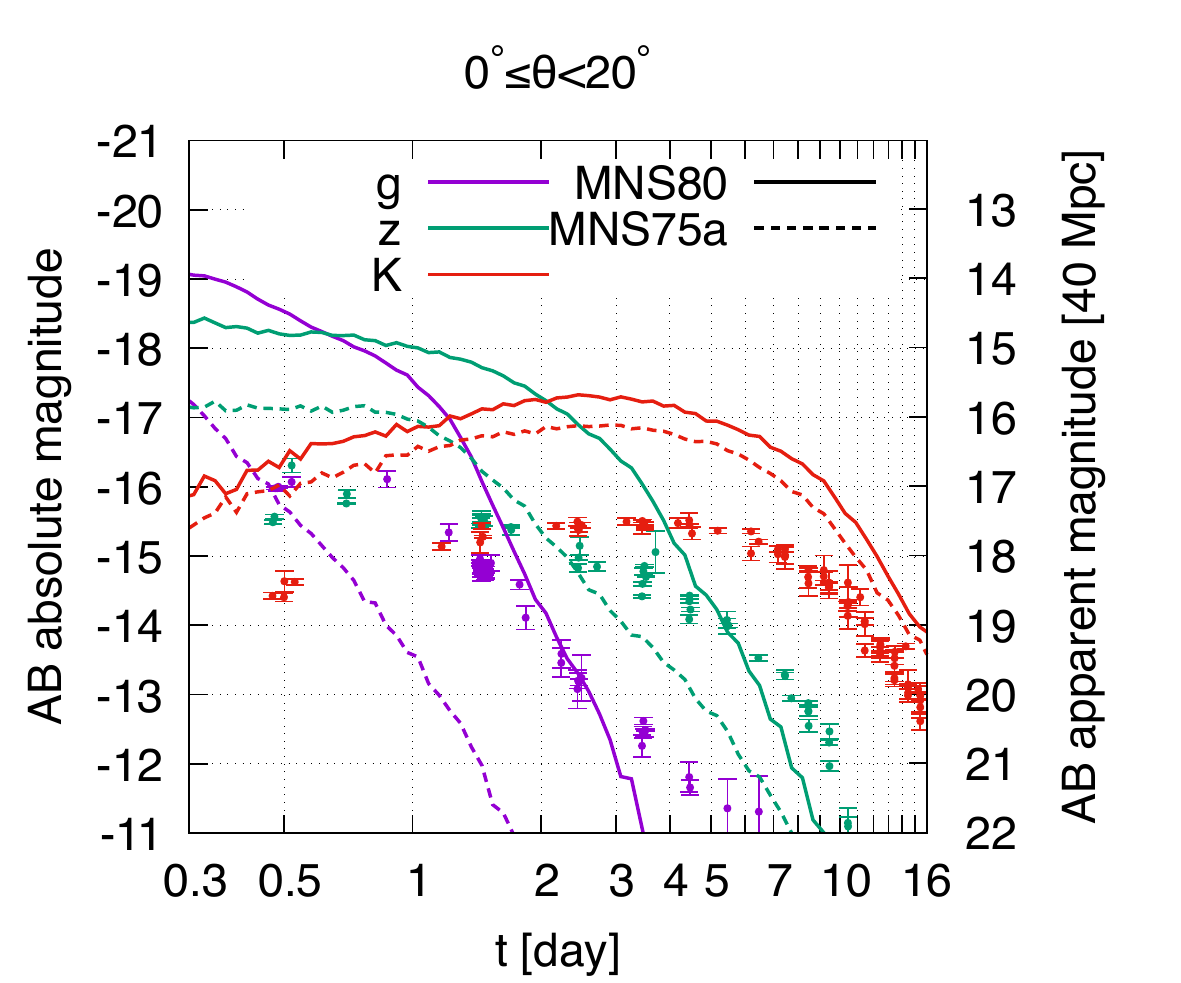}
 	 \includegraphics[width=.5\linewidth]{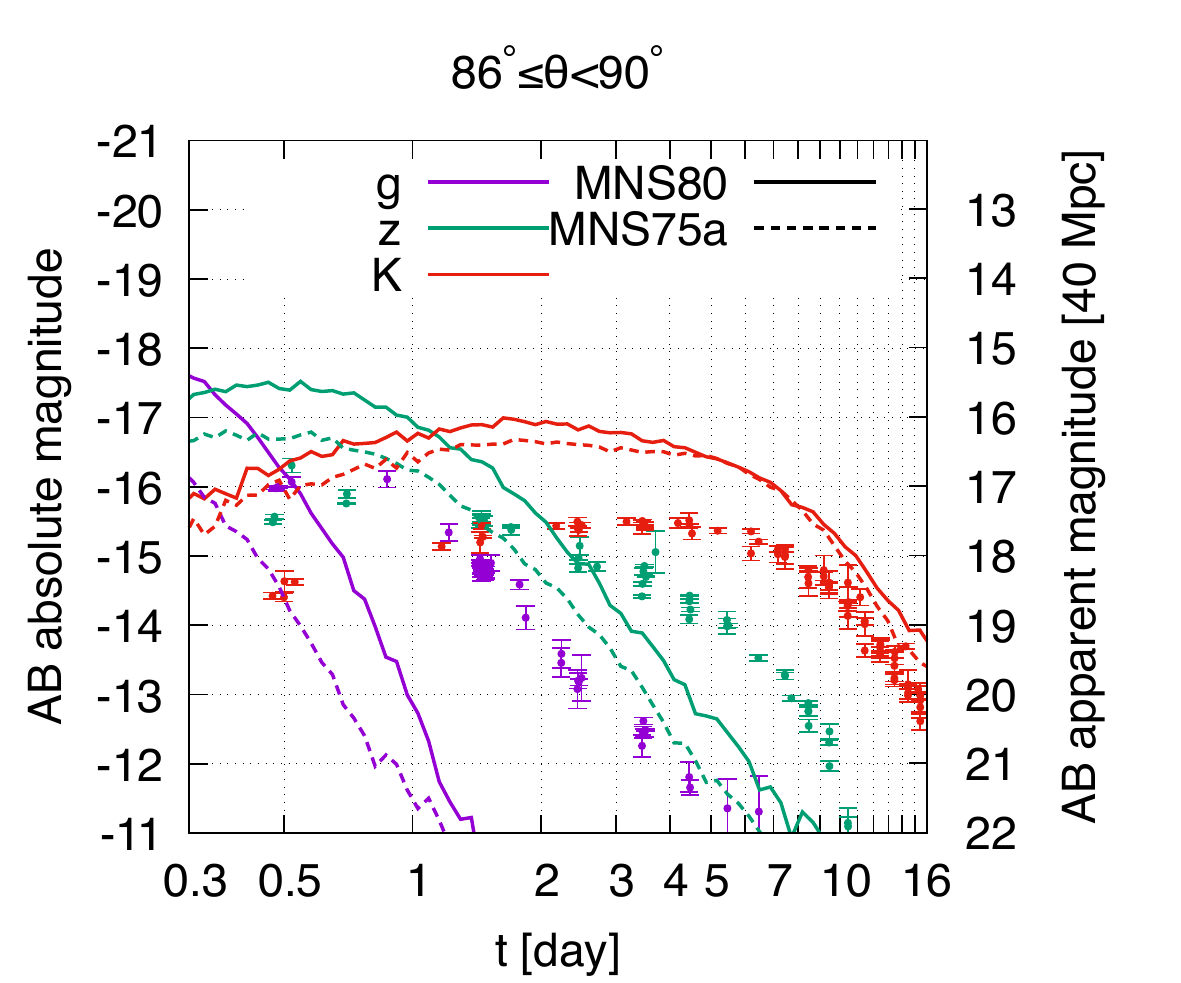}
 	 \caption{Bolometric luminosity and {\it gzK}-band light curves for model MNS80 in the same formats as in Figure~\ref{fig:bol} and Figure~\ref{fig:mag}. The light curves for model MNS75a are also plotted as references. The data points denote the bolometric luminosity and broad-band magnitude data of GW170817 taken from ~\cite{Waxman:2017sqv} and ~\cite{Villar:2017wcc}, respectively.} 
	 \label{fig:mag80}
\end{figure*}

\begin{figure}
 	 \includegraphics[width=\linewidth]{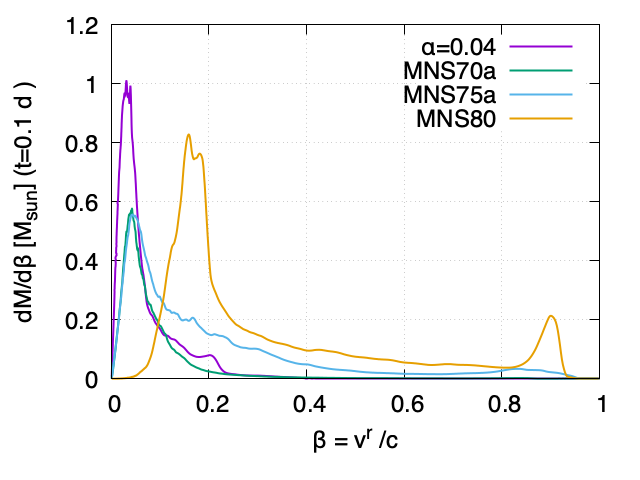}
	 \caption{The ejecta mass distribution at $t=0.1\,{\rm d}$ as a function of the radial velocity ($\beta=v^r/c$). }
	 \label{fig:dmdv}
\end{figure}

Next, we pay attention to model MNS80, in which the MHD effects are most significantly found among the models studied in this work.  Figure~\ref{fig:mag80} shows the bolometric luminosity and {\it gzK}-band light curves for this model in the same formats as in Figure~\ref{fig:bol} and Figure~\ref{fig:mag}, respectively, except for the enlarged plot range of the vertical axes. The kilonova emission for model MNS80 is always brighter than that for MNS75a. In particular, for $t\lesssim 1{\rm d}$, the total luminosity and isotropic luminosity observed from the polar direction ($\leq 20^\circ$) for model MNS80 are larger by factors of $3$ and $5$, respectively, than those for model MNS75a. We note that these differences are much larger than the difference in the total heating rate (see the top-left panel of Figure~\ref{fig:mag80}), of which the difference is primarily due to the difference in the total ejecta mass (see Table~\ref{tb:model}). On the other hand, for $t\gtrsim 1\,{\rm d}$, the total bolometric luminosity and isotropic luminosity for the viewing angle larger than $40^\circ$ for model MNS80 are larger than those for model MNS75a only by a factor of $1.5$--$2$, which approximately corresponds to the difference in the total ejecta mass. This implies that the difference in the luminosity in the late phase is primarily due to the difference in the total ejecta mass.

The bright emission in the early phase of model MNS80 is partially due to the high bulk velocity of the ejecta. Figure~\ref{fig:dmdv} shows the ejecta mass distribution at $t=0.1\,{\rm d}$ as a function of the radial velocity ($\beta=v^r/c$). While the ejecta matter distribution is mostly concentrated in the velocity smaller than $\approx 0.2\,c$ for models MNS75a, MNS70, and the viscous model, the ejecta matter is distributed in a higher-velocity region for model MNS80; the ejecta matter distribution has the peaks at the velocity of $\approx 0.2\,c$ and $\approx 0.9\,c$. The high bulk velocity of ejecta for this model helps photons to diffuse out more rapidly from the high-velocity edge of the ejecta, and hence, enhances the luminosity.

The absence of the first {\it r}-process peak elements in the polar high-velocity region is the other reason for the bright polar emission of model MNS80. As we already pointed out, some of the first-peak {\it r}-process elements (Y and Zr) have a significant contribution to the opacity in the optical wavelength. We find that the total Y $+$ Zr mass in the polar region of $\theta \leq30^\circ$ with $r/ct\leq0.3$ is by an order of magnitude smaller for model MNS80 than for models MNS70a and MNS75a. As a consequence, the {\it gz}-band light curves for model MNS80 are significantly brighter than those for other models, while the {\it K}-band light curve does not show an appreciable difference among the models.

The kilonova emission for model MNS80 is also much brighter than that of GW170817 with the hypothetical distance of 40 Mpc. In particular, the {\it gz}-band magnitudes of model MNS80 for $t\le1\,{\rm d}$ are by 2 mag brighter for the polar view ($\leq 20^\circ$), although they decline much rapidly after $t\approx1\,{\rm d}$. As is the case for model MNS75a, this suggests that the MHD effect might not play an important role in the post-merger phase of GW170817. On the other hand, if a BNS merger which results in a long-lived MNS with the MHD effect as significant as for model MNS80 occurs, our result indicates that the kilonova can be as bright as in GW170817 even at the distance of 120\,Mpc.


\subsection{possible non-LTE effect}\label{sec:nonLTE}

\begin{figure*}
 	 \includegraphics[width=\linewidth]{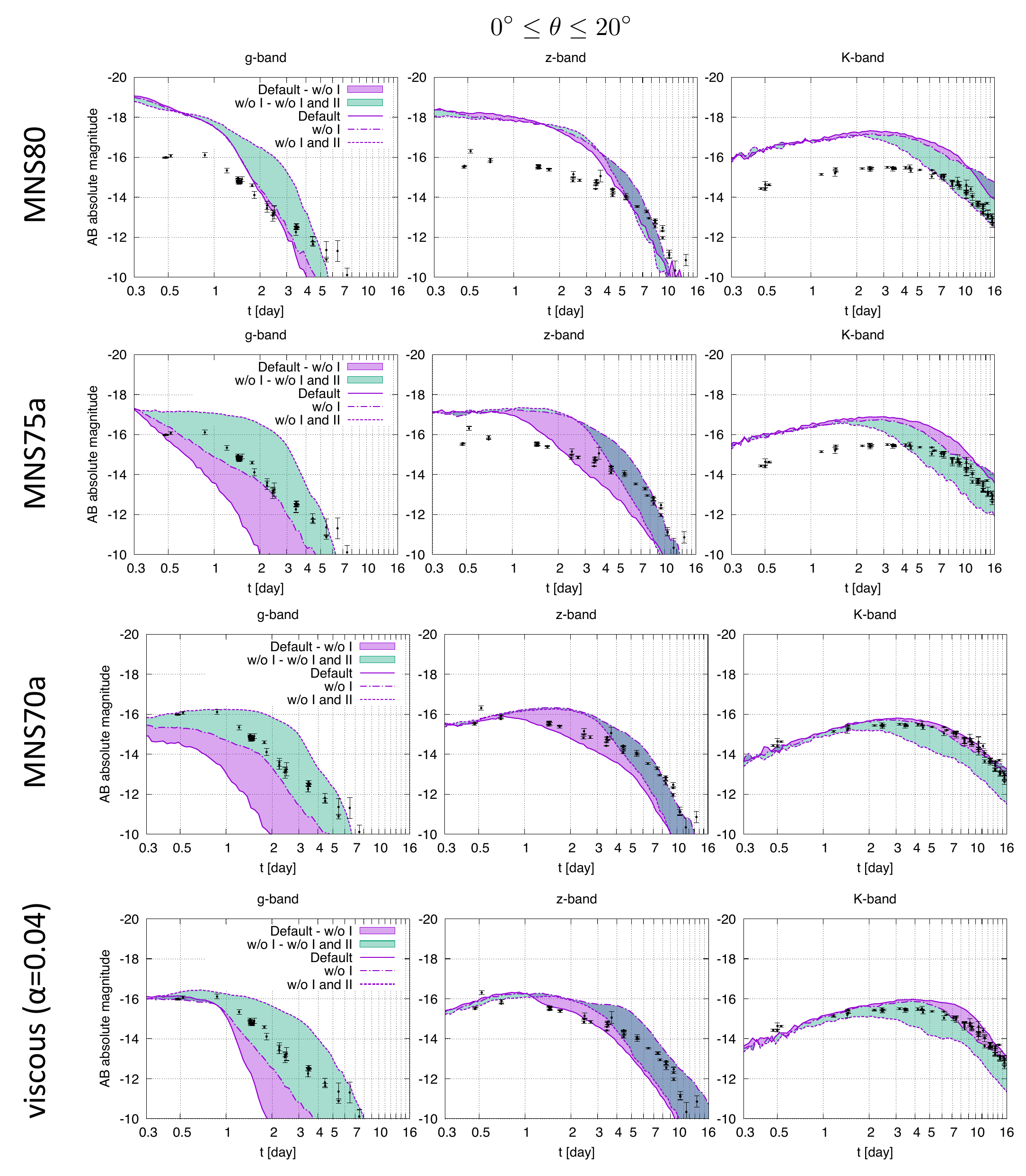}
 	 \caption{Comparison of the {\it gzK}-band light curves observed from $0^\circ\leq\theta\leq20^\circ$ for models in which the opacity contributions from the neutral (``w/o I"; dash-dot curves) or both neutral and first ionized atoms (``w/o I and II"; dotted curves) are switched off. The light curves shown in the solid curves (``Default") are the results with the default setting and are the same as in Figure~\ref{fig:mag}. The left, middle, and right panels denote the {\it g},  {\it z}, and {\it K}-band light curves, respectively. The first, second, third, and fourth panels from the top denote the models MNS80, MNS75a, MNS70a, and the viscous model ($\alpha=0.04$),  respectively. The data points denote the observation data of GW170817 taken from~\cite{Villar:2017wcc} for a hypothetical distance to the source of 40\,Mpc.}
	 \label{fig:nonLTE}
\end{figure*}

For the later phase of the kilonova emission (for $t\gtrsim1\,{\rm d}$), the condition of LTE, which we assumed in the present radiative transfer simulations, could break down for the low density region in which the ionization of the atoms by the radioactive rays becomes more significant than the recombination of ions~\citep{Hotokezaka:2021ofe,Kawaguchi:2020vbf}. In fact, the importance for taking the non-LTE effect in the excitation/ionization population into account is well known for supernovae radiative transfer simulations~\citep[e.g.,][]{Boyle:2016zcr}. 

\cite{Hotokezaka:2021ofe} solved the ionization and electron temperature evolution in the kilonova nebular phase using a model in which the atomic properties of the ejecta are represented by Nd ions. They found that the population of the neutral and first ionization atoms are significantly suppressed in the low density region because of the high efficiency of radioactive ionization  due to  $r$-process nuclei. \cite{Pognan2022MNRAS} also obtained similar results on the ionization evolution in the nebular phase.
These results indicate that the non-LTE effects may suppress the neutral and first ionized ions in the outer part of the ejecta even in the earlier phases. In fact,~\cite{Pognan:2022pix} show that the non-LTE effect can significantly modify the ejecta opacity particularly in the low density region even at a few days after the merger.
Because computing the ionization/excitation population is challenging due to its computational complexity and lack of the atomic data for the $r$-process elements (see~\citealt{Hotokezaka:2021ofe,Pognan2022MNRAS}), we here provide qualitative estimates for the impacts of the non-LTE effects on the kilonova light curves by removing the neutral or both neutral and first ionized atoms. For this purpose, we perform the radiative transfer simulations with a hypothetical setup in which the neutral or both neutral and the first ionized atoms obtained by solving the Saha's equations are artificially forced to be ionized to the first or the second ionization states, respectively. Note that this prescription is applied to whole ejecta including a high density regions for simplicity. 

Figure~\ref{fig:nonLTE} shows the {\it gzK}-band light curves observed from $0^\circ\leq\theta\leq20^\circ$ for models in which the opacity contributions from the neutral (``w/o I"; dashed curves) or both neutral and first ionized atoms (``w/o I and II"; dotted curves) are switched off. The light curves with the default setting shown in Figure~\ref{fig:mag} are also plotted with the solid curves in Figure~\ref{fig:nonLTE} as the references. 
By suppressing the opacity contribution from the neutral or the first ionized atoms, the brighter and fainter emission is realized at the time of the peak magnitudes in the {\it gz}-bands and {\it K}-band, respectively. Furthermore, the time of the peak magnitudes moves to later and earlier epochs in the {\it gz}-bands and {\it K}-band, respectively. The effect of suppressing the opacity contribution from the neutral or the first ionized atoms is more significant for the emission in the shorter wavelengths and the effect on the {\it g}-band is significant even from $t\approx0.5\,{\rm d}$. On the other hand, the {\it zK}-band light curves for $t\lesssim1\,{\rm d}$ is less affected by the suppression of the opacity contribution from the neutral or the first ionized atoms.

The extinction of the polar optical emission, which is seen for models MNS75a and MNS70a (see Figure~\ref{fig:mag}), is suppressed by switching off the opacity contribution from the neutral and the first ionized atoms. The {\it g}-band light curves for both model MNS70a and viscous model become in better agreement with the observed data points of GW170817 if we suppose that the light curves could vary within the ranges up to those calculated without the opacity contribution from both neutral and the first ionized atoms. On the other hand, Figure~\ref{fig:nonLTE} indicates that the {\it K}-band emission for $t\gtrsim2\,{\rm d}$ can become fainter than the observation by the non-LTE effect, though it is still in the range of the uncertainty. This might imply that a different type of BNS merger which produces brighter NIR photons in the kilonova, such as a BNS with unequal mass NSs or a BNS of which remnant MNS collapses to a black hole within a short time scale ($\lesssim 100$ ms), can be more consistent with the observed data of GW170817.
 
The light curves for models MNS80 and MNS75a show different features from those observed in GW170817 even if the opacity contribution from the neutral or the first ionized atoms is suppressed: though the difference in the {\it g}-band emission from the observation can be less pronounced by the non-LTE effect, the {\it z} and {\it K}-band emission is still always brighter than the data points by more than 1 magnitude for $t\lesssim10\,{\rm d}$. These results support our hypothesis that a long-lived MNS associated with a strong global magnetic field is unlikely to be formed in GW170817.

Comparing with the other models, the effect by switching off the opacity contribution from the neutral and the first ionized atoms is relatively minor for model MNS80, especially for the {\it gz}-band light curves. This reflects the fact that Y and Zr in the polar region is less abundant for model MNS80 than the other models, and the suppression of the optical light curves is not significant in the first place. 

Figure~\ref{fig:nonLTE} shows that both brightness and color of the kilonova emission can be significantly modified if the non-LTE effect plays an important role for determining the ionization population of the ejecta. Beside the non-LTE effect, we note that the uncertainty in the abundance distribution (and thus radioactive heating rate) could also be the source of the systematic error for the light curves because not only the emissivity but also the ejecta temperature distribution, which is responsible for the ionization structure~\citep{Barnes:2020nfi}, are modified by it. Hence, the systematic study employing different nucleosynthesis models will be needed for quantitatively understanding the uncertainty in the light curve.

\section{Conclusions and Discussion}\label{sec:dis}

Our results suggest that  the presence of the bright synchrotron flare would be an indicator for the presence of long-lived MNSs with significant amplification of the global magnetic field, if we can distinguish it from the jet afterglow. Such synchrotron emission can be observed even for a far distance for which gravitational waves are not detectable. For example the radio afterglow of the MHD models with $\sigma_{\rm c}\ge3\times10^{7}\,{\rm s^{-1}}$ and with an ISM density of $>10^{-3}\,{\rm cm^{-3}}$ at a distance of $200$ Mpc can be identified with untargeted surveys~\citep[see, e.g.,][]{Hotokezaka2016ApJ,Dobie2021MNRAS}. In fact, \cite{Dobie2022MNRAS} demonstrated that an untargeted search  with ASKAP can identify radio transients on time scales of a few days to a year with a flux level down to $\sim 200\,{\rm \mu Jy}$ if the localization area is reasonably small $\sim 30\,{\rm deg^2}$. They found one radio transient in the localization area of GW190814, which is likely unrelated to the merger event.

The ASKAP untargeted search for GW190814 provides an interesting upper limit on the surface density of radio transients above $170\,{\rm \mu Jy}$ varying on time scales of days to a year as $<0.013\,{\rm deg^{-2}}$~\citep{Dobie2022MNRAS}. If we suppose that a fraction $f$ of BNS mergers with a rate of $300\,{\rm Gpc^{-1}\,yr^{-1}}$ produces a radio transient similar to model MNS80 with $n=10^{-3}\,{\rm cm^{-3}}$ (see the middle panel of figure \ref{fig:radio}) the surface density of such transient is expected to be $\sim 0.03f\,{\rm deg^{-2}}(S_{\nu}/170\,{\rm \mu Jy})^{-3/2}$. This observed upper limit implies $f\alt 0.3$. Thus, BNS mergers of which remnant MNSs survive for a long period could be minority, if significant MHD effects are always induced in the remnant MNSs. This is consistent with our latest finding~\citep{Fujibayashi:2020dvr}, which shows that in the presence of long-lived MNSs, relatively light $r$-process elements should be over-produced and the entire solar-abundance pattern of the $r$-process elements, which is believed to be the universal, cannot be reproduced. 

On the other hand, we emphasize that not necessarily all the BNS mergers produce ejecta with the solar one, though it should be not a major type of BNSs if the solar $r$-process-abundance pattern is universal. For example, if a metal-poor star for which the elemental abundance pattern is different from the solar one, it might also be explained by the $r$-process nucleosynthesis in some rare type of a BNS merger, and we do not always have to consider that it was proceeded in a source different from the BNS merger. We also emphasize that a bright long-lasting radio transient has not been found after short GRBs even though the extensive follow-up observations have been conducted  \citep{Metzger2014MNRAS,Horesh2016ApJ,Fong2016ApJ,Klose2019ApJ,Schroeder2020ApJ}. This implies that BNS mergers which result in long-lived MNSs with significant amplification of the global magnetic field (i.e., magnetars) might not be the central engine of short GRBs. 

In the presence of a long-lived MNS with a strong global magnetic-field, the kilonova emission will also show distinct features from the cases that the magnetic-field amplification and its retaining are not significant or the MNS collapses to a BH in a short time scale ($\lesssim 100\,{\rm ms}$); the optical emission which is bright for $\alt 1$\,d but becomes steeply weak in a short time scale of a few days and the NIR emission which is always brighter by more than 1 magnitude than those observed in GW170817 for $t\lesssim10\,{\rm d}$. While we should note that the presence of such features, particularly in the optical wavelength, depends on how strongly non-LTE effects play a role (see Sec.~\ref{sec:nonLTE}), the significant amplification of the global magnetic field can be examined by observing such features in the early epoch. For this purpose, the rapid kilonova search with a high cadence is important.  

The bolometric luminosity for the viscous model and MNS70a (except that observed from $\theta\le30^\circ$) agrees fairly well with the observed data of GW170817. The peak magnitudes in the {\it grizJHK} bands for these models are also in good agreement with the observation of GW170817. The disagreements of the light curves after the time of the peak magnitudes will be much less pronounced if we consider the possible uncertainty of the light curve prediction due to the non-LTE effects. On the other hand, we showed that the kilonova light curves for the MHD models with $\sigma_{\rm c}\ge3\times 10^7\,{\rm s^{-1}}$ will exhibit different features from GW170817. This suggests that the MHD effect might not play an important role in the post-merger phase of GW170817. Since the significance of the MHD effect depends strongly on the parameters of the phenomenological dynamo model, it is not clear whether, and if so, at which time the remnant MNS collapsed to a BH. Thus, investigation for the realistic value for the phenomenological dynamo parameters as well as the quantitative evaluation of the non-LTE effects will be important tasks to interpret the systems of GW170817 and the future events. 

\acknowledgments 
KK thanks Smaranika Banerjee, Nanae Domoto, and Masaomi Tanaka for the valuable discussions. Numerical computation was performed on Yukawa21 at Yukawa Institute for Theoretical Physics, Kyoto University and the Sakura, Cobra, Raven clusters at Max Planck Computing and Data Facility. This work was supported by Grant-in-Aid for Scientific Research (JP20H00158, JP21K13912) of JSPS/MEXT and Japan Society for the Promotion of Science (JSPS) Early-Career Scientists Grant Number 20K14513.


\bibliographystyle{apj}
\bibliography{ref}

\appendix
\section{Treatment of radioactive heating}\label{app:heat}
In this paper we update the treatment of radioactive heating in the HD simulation from the previous study, based on the formulation of~\cite{2016MNRAS.456.1320T,Uchida:2017qwn}. In the previous study~\citep{Kawaguchi:2020vbf}, the rest-mass density, $\rho_*$, was evolved as a conserved quantity in solving the continuity equation. However, strictly speaking, using $\rho_*$ as the conserved quantity is incorrect in the presence of the radioactive reaction because the rest-mass of baryons can be changed due to the change of the nuclear binding energy. Thus, in this work, we define ${\hat \rho}=m_{\rm u} n_{\rm B}$ where $m_u$ is the atomic mass unit and $n_{\rm B}$ is the baryon number density, and define ${\hat \rho}_*={\hat \rho} u^t \sqrt{g}$ as a conserved quantity which follows the continuity equation without a source term; that is,
\begin{align}
	\partial_t{\hat \rho}_*+\partial_i\left({\hat \rho}_*v^i\right)=0,
\end{align}
where $v^i:=u^i/u^t$.  

The release of the nuclear binding energy by the radioactive decay is described by the change in the mass per baryon, $m_{\rm b}$: We define the mass excess of the baryon by
\begin{align}
	\delta_{\rm exc}:=\frac{m_{\rm b}-m_{\rm u}}{m_{\rm u}}.
\end{align}
Using $\delta_{\rm exc}$, the rest-mass density is given by $\rho=(1+\delta_{\rm exc}){\hat \rho}$. The enthalpy per baryon particle (weighted by $1/m_{\rm u}$), ${\hat h}$ is then written as
\begin{align}
	{\hat h}=\left(1+\delta_{\rm exc}\right)c^2+{\hat \epsilon}+\frac{P}{{\hat \rho}},
\end{align} 
with $\hat\epsilon$ being the internal energy per baryon particle (weighted by $1/m_{\rm u}$). Note that ${\hat h}$ is related to the specific enthalpy (enthalpy per rest mass), $h$, defined in the previous paper with $\rho h={\hat \rho} {\hat h}$. 
Then, the energy-momentum tensor is written as
\begin{align}
	T_{\mu\nu}={\hat \rho} {\hat h}u_\mu u_\nu+Pg_{\mu\nu}.
\end{align}

Considering that a fraction of the energy release by the radioactive decay is carried and lost from ejecta by the neutrino emission, the energy-momentum conservation is written as
\begin{align}
	\nabla_\mu T^{\mu\alpha}=-{\hat \rho}{\dot q}_{(\nu)}u^\alpha(\tau),
\end{align} 
where ${\dot q}_{(\nu)}$ denotes the neutrino energy deposition rate per baryon particle  (weighted by $1/m_{\rm u}$) due to the radioactive decay and $\tau$ denotes the proper time (affine parameter) in the fluid rest frame. The time evolution of the mass excess in the fluid rest frame is connected to the radioactive-heating rate per baryon particle (weighted by $1/m_{\rm u}$), ${\dot q}_{\rm tot}$, with
\begin{align}
	\frac{d\delta_{\rm exc}}{d\tau}=-\frac{1}{c^2}{\dot q}_{\rm tot}(\tau).
\end{align}
We can rewrite this equation into a form which is similar to the conservation form of the hydrodynamics; 
\begin{align}
	\nabla_\mu\left({\hat \rho} \delta_{\rm exc}u^\mu\right)=-\frac{1}{c^2}{\hat \rho}{\dot q}_{\rm tot}(\tau).
\end{align}
To evaluate ${\dot q}_{(\nu)}$ and ${\dot q}_{\rm tot}$ at each time step, the proper time in the fluid rest frame is needed. For this purpose, we also determine the proper time of each fluid element by solving
\begin{align}
	\nabla_\mu\left({\hat \rho} \tau u^\mu\right)={\hat \rho},
\end{align}
which is equivalent to solving $d\tau/d\tau=u^\mu\nabla_\mu\tau=1$.
\section{Numerical convergence of the ejecta kinetic energy distribution}\label{sec:conv}

\begin{figure*}
 	 \includegraphics[width=1.\linewidth]{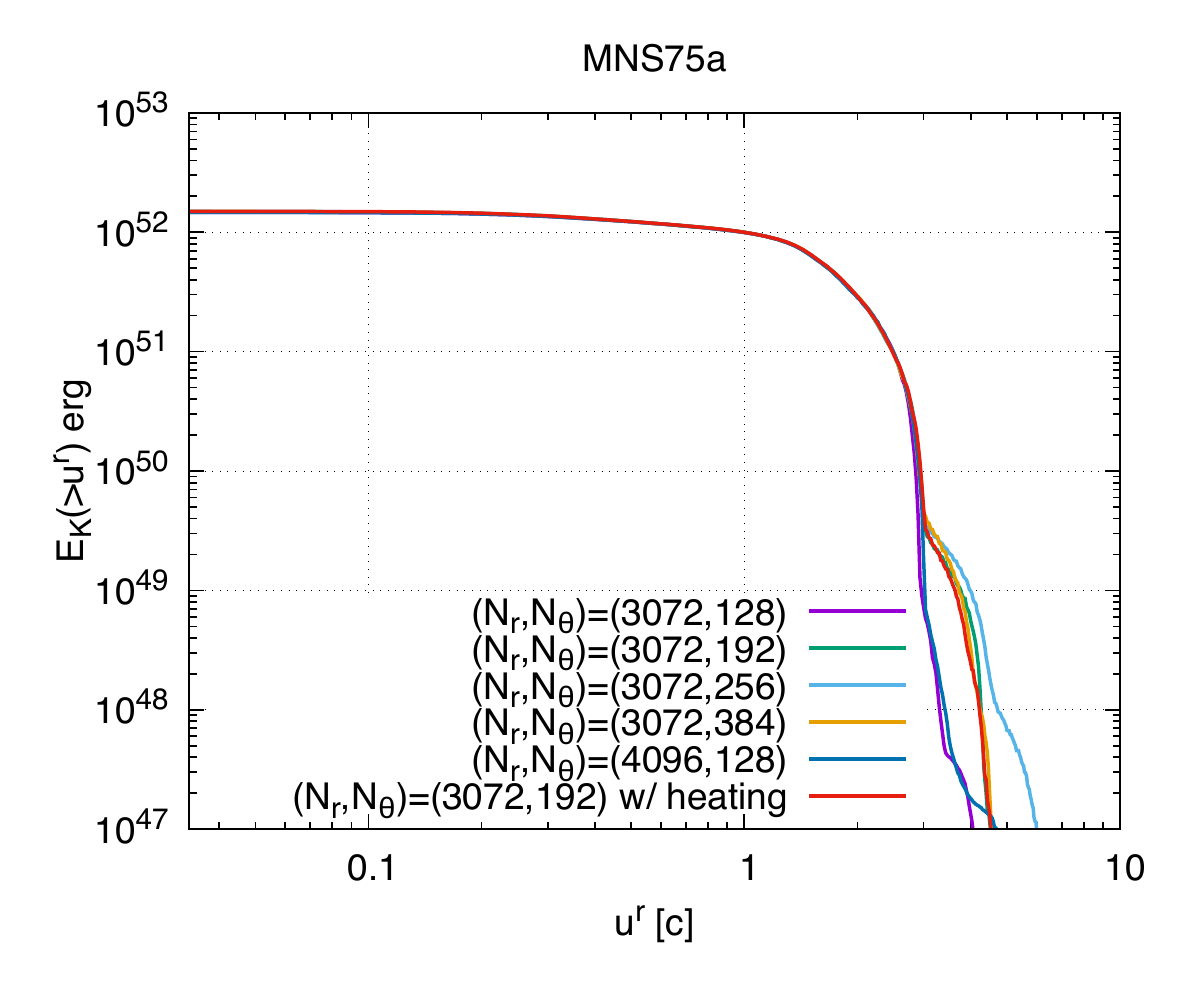}\
 	 \caption{Convergence study of the kinetic energy distribution for MNS75a. $N_r$ and $N_\theta$ denote the numbers of the grid points in $r$ and $\theta$ directions, respectively. The curves with the label ``w/ heating" denotes the result in which radioactive heating is taken into account.}
	 \label{fig:keconv}
\end{figure*}

Figure~\ref{fig:keconv} compares the kinetic energy distribution for MNS75a obtained by the HD simulations with various grid resolutions. We found that the model with $N_r=3072$ and $N_\theta=192$ is fairly sufficient to resolve the ejecta fast tail with $u^r/c\lesssim 5$, where $N_r$ and $N_\theta$ denote the numbers of the grid points in $r$ and $\theta$ directions, respectively.

We also confirmed that the light curves of the kinetic energy distribution is not affected by the presence of the radioactive heating terms in the HD simulations (see the curves with ``w/ heating" in Figure~\ref{fig:keconv}). This can be understood by the fact that the energy deposition due to radioactive heating is negligible compared to the kinetic energy for the fast tail. 

\end{document}